\documentclass[a4paper,12pt]{article}
\usepackage{graphicx}

\usepackage{caption}
\usepackage{bm}
\newcommand{\Vec}[1]{\mbox{\boldmath$#1$}}

\newcommand{\bk}{{\Vec k}}

\begin{document}
\begin{flushright}
To be published in {\it Contemporary Physics}
\end{flushright}

\begin{center}
{\large\textbf{Flat bands in condensed-matter systems \\
--- perspective for magnetism and superconductivity
}\par
\ \\ 
Hideo Aoki}
\par
\ \\
{\it Department of Physics, University of Tokyo, Hongo,
Tokyo 113-0033, Japan}

ORCID identifier:  http://orcid.org/0000-0002-7332-9355

\end{center}
\par
\ \\

{\bf Abstract}  

There is a recent upsurge of interests in 
flat bands in condensed-matter systems and the consequences for magnetism and superconductivity.  This article highlights the physics, where 
peculiar quantum-mechanical mechanisms for the physical properties 
such as flatband ferromagnetism and flatband superconductivity 
that arise when the band is not trivially flat but has a strange 
Hilbert space with non-orthogonalisable Wannier states, 
which goes far beyond just the diverging density of states.  
Peculiar wavefunctions come from a quantum-mechanical 
interference and entanglement.  
Interesting phenomena 
become even remarkable when many-body interactions are 
introduced, culminating in flatband superconductivity 
as well as flatband ferromagnetism.  Flatband physics harbours a very wide range physics 
indeed, extending to non-equilibrium physics in laser illumination, 
where 
Floquet states for topologcial superconductivity is promoted in 
flatbands.  
While these are theorecially curious, possible candidates for 
the flatband materials are beginning to emerge, which 
is also described.  These provide a wide and promising outlook.


{\bf Contact}: Hideo Aoki email: aoki@phys.s.u-tokyo.ac.jp

\section{What is the flat band?}

\subsection{Flatbands in a nutshell}

The word ``flat" appears in physics in diverse contexts.  Formerly, 
we often heard about 
a flatland, which is meant to signify spatially two-dimensional (2D) 
systems.  Typically, the 2D electron system confined on a plane is 
an arena for various exotic phenomena, where 
the quantum Hall effect is a prime example.  
There, we were talking about 
a flatness in real space.  In a totally different avenue, 
we can think of a flatness in a momentum ($k$) space, 
and the flatband refers to that.  
However, we have to caution from the 
outset:  a band that has a flat dispersion in $k$ space would imply, 
trivially, that an electron cannot hop between the atomic 
sites in a crystal, and this case is called 
the atomic limit.  However, the flatband which 
is attracting recent interest and the subject of the 
present article is a totally different class of 
flatbands, where 
dispersionless bands arise despite a nonzero hopping, 
with a most familiar realisation occurring in 
kagome lattice.  Mathematically, the flatband has not just 
zero bandwidth, but harbours quite an 
anomalous situation where the Wannier functions, 
usually definable as an orthonormal basis, do not exist on the 
flatband due to quantum interference, or frustration in 
wavefunctions.  

This brings about multitude of 
unusual properties in condensed-matter physics, and 
the flatband physics  and the systems 
exhibiting them have a very wide spectrum indeed.  
An overview of the topics I shall expound in the present article 
is summarised in {\bf Fig.1}.
This spanns over magnetism, 
superconductivity, and topology.  In a wider perspective, 
the horizon expands even further if we go from equilibrium 
systems to non-equilibrium cases.  The purpose of the 
present article is to give basics and perspectives 
over these speactra, thereby emphasising that the key factor of the 
flatband is peculiar {\it quantum interference} and {\it quantum 
geometry}.   
Nonequilibrium physics, 
which in general hosts versatile quantum states that are 
unthinkable in equilibrium, becomes even 
more interesting for flatbands in laser lights, so a section is devoted 
to that.  
Throughout the article, we shall focus on the physical concepts and 
materials science aspects rather than technical details.

\begin{figure}[ht]
\begin{center}
  \includegraphics[width=1.0\textwidth]{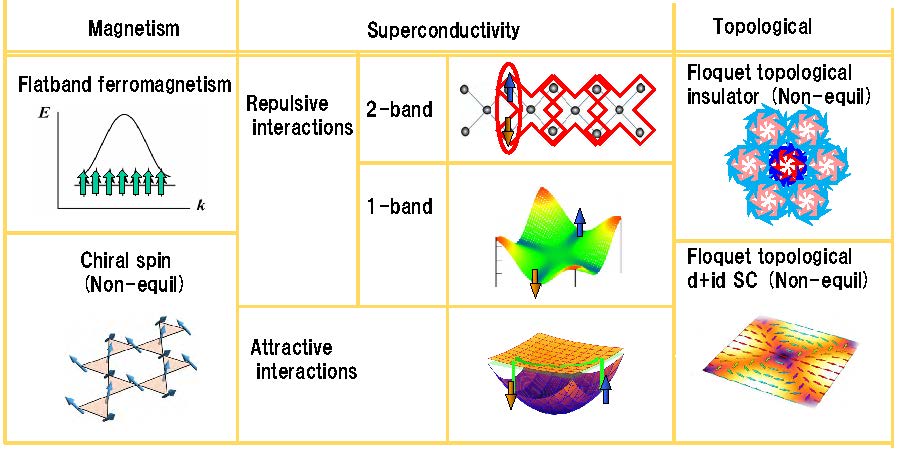}
\caption{
The topics covered in the present article for 
various phenomena with various settings. 
}
\label{fig_classificationTable_magnetism}     
\end{center}  
\end{figure}

\subsection{Some background}

Before plunging into the flatband physics, 
let us just briefly give a reminder of the key, relevant background of 
e.g. what physics is expected in 
flatband systems as compared with the ordinary systems, 
in order to set the scene.  

First, the band structure calculations come in two flavours, where 
one is for the nearly-free electron systems (e.g. 
alkali metals), while the other 
is for the electrons more strongly bound to each atom (typically 
transition metal compounds, organic solids, etc) where 
the band structure reflects the lattice structure more 
strongly, with 
the standard way being to use the tight-binding model.  
Since the flatbands inherently come from the lattice structures, 
we usually use the tight-binding models with the tight-binding approximation.  

For many-body effects, weakly-interacting systems can be dealt with 
mean-field and perturbative methods, while more sophisticated methods are 
required to describe the electron correlation arising from the 
interaction.  This especially applies to flatband cases, since the 
flatness of the band infensifies the correlation effects.

For magnetism, for weakly-interacting systems, 
we usually have paramagnetism and Landau's diamagnetism.  
For correlated electron systems as examplified by transition metals 
and their compounds, strong-coupling methods are required, 
especially for ferromagnetism and antiferromagnetism.  

For superconductivity, for weakly-interacting systems, 
BCS theory is the standard starting point, 
which captures the superconductivity for ordinary, 
low-Tc superconductors.  The electron-phonon 
coupling is the source of the effective attraction 
in conventional superconductors, and, for strong coupling, 
more sophisticated methods should be 
applied.  We have quite a different picture for 
strong electron-electron interactions, which are 
at the core of the high-Tc superconductors exemplified 
by the cuprates (copper compounds).  

Topological states have now become one of the most 
active fields in the condensed matter physics.  
This also requires sophisticated methods, because 
the topological states are qualitatively unlike the 
ordinary quantum phases which are basically 
understandable in term of the order parameters.  

So the questions we expound in the present article are:  
What special properties arise in the flatband 
systems in terms of their band structure (mainly with 
the tight-binding model), their magnetism (mainly 
the flatband ferromagnetism), superconductivity 
peculiarly enhanced in flatbands (flatband superconductivity), and topological 
properties (here mainly for non-equilibrium situations).

\subsection{Lieb, Mielke and Tasaki models and anomalous Wannier states}

Now, let us start with the definition of the flatband.  Before the 
advent of the flatband physics, a flatband merely meant 
the atomic limit and is uninteresting.  
Then came the flatbands according to Elliott Lieb back in the 
1980s, which was followed by 
the model due to Andreas Mielke, and due to Hal 
Tasaki\cite{Lieb:1989,Mielke:1991,Tasaki:1992}.  
{\bf Figure 2} 
displays them for two-dimensional tight-binding models.  
A Lieb lattice is obtained from an ordinary Bravais lattice 
by adding an extra site in the 
middle of each bond\cite{Lieb:1989}.  This makes 
the number, $N_A$, of A sublattice sites 
different from that of B sublattice sites, $N_B$, 
for bipartite lattices.  
We can then see, by counting the rank of the Hamiltonian matrix in the tight-binding model as Lieb argues, that the matrix has to have at least 
$|N_A-N_B|$ zero eigenvalues, which correspond to the flatband(s).

\begin{figure}[ht]
\begin{center}
  \includegraphics[width=1.0\textwidth]{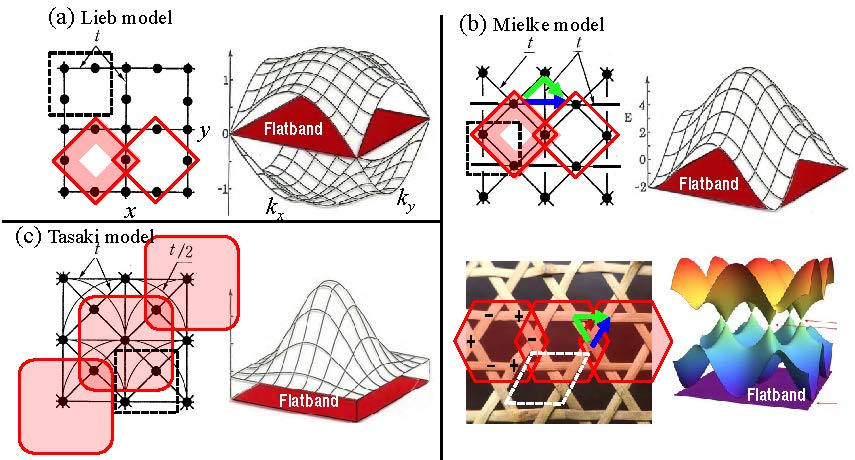}
\caption{
Lieb (a), Mielke (b) and Tasaki (c) flatband models.  In each panel, the lattice structure is displayed on the left with red enclosures representing 
(overlapping) Wannier functions, and the band dispersion on the right.  Dashed lines enclose unit cells.  In (b), a tetragonal (upper) and kagome (lower; line-graph constructed from honeycomb) 
realisations are displayed, where the latter is simpler in that no two bonds cross each other in a top view.  Two interfering paths as an electron hops from a site to the next are shown in blue and green arrows.  Right bottom figure is from JPS Hot Topics 2, 030 (2022).
}
\label{fig_LiebMielkeTasaki}     
\end{center}  
\end{figure}

A Mielke model (Fig.2 
(b)) 
is obtained when units (tetrahedra 
or triangles in the examples in this figure) 
are connected by sharing apices\cite{Mielke:1991}.  The emergence of 
flatbands is elucidated in terms of the line-graph 
and molecular-orbital pictures as we shall describe below.  
In a Tasaki model (Fig.2
(c)), the flatband is separated from a
dispersive one with a band gap\cite{Tasaki:1992}.  
These classes of models can be systematically constructed for any spatial 
dimensions.  For instance, a typical 3D Mielke model is 
the pyrochlore lattice, which is a 3D version of kagome\cite{aokiJPCM04}.

What is special in these classes of models is that the wavefunctions 
on flatbands are anomalous 
in that the Wannier basis functions, strangely enough, 
{\it cannot be orthogonalised}.  Namely, in an introductory 
solid-state physics, a standard procedure for treating the 
electrons in 
a crystal (a spatially periodic system) is to first 
construct the Bloch wavefunctions in the momentum space, then Fourier-transform the basis 
to have the Wannier basis in real space.  While this may seem always feasible, 
that is violated if we have the flatband systems {\it \`{a} la} Lieb, Mielke and 
Tasaki, where the smallest possible Wannier functions have to 
overlap with each other as displayed in Fig.2.  
If we force them to be orthogonal with the Gram-Schmidt 
orthonormalisation, the procedure would expand the functions.  
The quantum states are in fact shown to be 
stronlgy {\it entangled} as we shall see.

One way to see 
why the band is flat despite the nonzero hopping is to realise 
that, when you go from a site to a neighbour, 
there are multiple paths as indicated in Fig.2
(b) 
for Mielke models.  There is a quantum mechanical interference 
between the paths, which works destructively in the 
flatbands.  In this sense the flatband comes from a 
kind of frustration, and this causes the unorthogonalisable Wannier states.  
This is at the 
core of the flatband ferromagnetism, 
and of the topological superconductivity in flatbands as well. 

Mathematically, there is now an intensive body of works 
on the orthogonalisability of 
Wannier functions and their sizes (called Wannier spread).  
We can itemise them as:

(i) For flatbands that 
satisfy the connectivity condition (see below), 
Wannier functions in the usual sense do not exist 
even when they are topologically-trivial\cite{MarzariVanderbilt97}. 

(ii) For topological  (with nonzero Chern numbers) but dispersive bands, 
Wannier states are undefinable\cite{Brouder07}. 

(iii) For topological flatbands, Wannier states are also absent.  
Historically, the quantum Hall effect (QHE) is the first 
system recognised as the topological system, where the phenomenon 
takes place on the Landau levels (a kind of flatband), 
and it has long been known that no Wannier states exist in QHE systems 
(see, e.g., Ref.\cite{aoki_SES2}).  

(iv) In condensed-matter physics in general, adiabatic arguments 
are often enlightening, where we discuss how the physical 
properties change as we adiabatically change some parameter 
that defines 
the system (see e.g. P.W. Anderson: {\it Basic Notions of Condensed Matter 
Physics} (Benjamin, 1984), Ch.3).  
We can then pose a question: is there an adiabatic route 
from a topological flatband having  an unorthogonalisable Wannier basis 
down to the atomic limit?  The answer is no --- 
topological systems are not adiabatically connected to the atomic limit\cite{poWatanabe17}.  
On topological flatbands, see also 
Ref.\cite{watanabe15,poWatanabe17} 
for what the authors call a fragile topology.  
A way to capture the topological flatbands is, 
as detailed in section `Topological flatbands and quantum-metric 
implications' below, to evoke the 
`quantum geometry', which may seem a fancy notion but 
is now recognised to be an important way of capturing 
the entire set of wavefunctions in a band\cite{loopstates}.  
The quantum geometry can also be used to 
look into a  relation of the flatband  with 
Landau levels\cite{Bohmyung20}.

As for the quantum interference,  
there is some difference between Lieb model and Mielke/ Tasaki models.  
One way to see this is 
to look at the electronic spectra when we apply external 
magnetic fields, $B$.  In general, when we apply a magnetic 
field to periodic systems such as crystals, 
very intricate (in fact fractal) energy spectra appear, which are called 
Hofstadter butterflies.  If we apply $B$ to 
flatband systems, the Hofstadter butterfly arises 
differently between Lieb and Mielke/Tasaki models as shown in 
{\bf Fig.3} 
\cite{aoki_Hofstadter}.  
Namely, in Lieb models, the flatband 
remains flat even in $B$ while the dispersive bands proliferate into 
butterflies.  In Mielke/Tasaki models, the flatbands 
proliferate into 
butterflies, too, where $B$ changes the interference in the 
hopping paths.  
The Lieb lattice is still affected by $B$, in that 
the compact localised state (i.e., overlapping yet smallest possible state) 
becomes an ``elongated ring state" in an external magnetic field.

\begin{figure}[ht]
\begin{center}
  \includegraphics[width=0.5\textwidth]{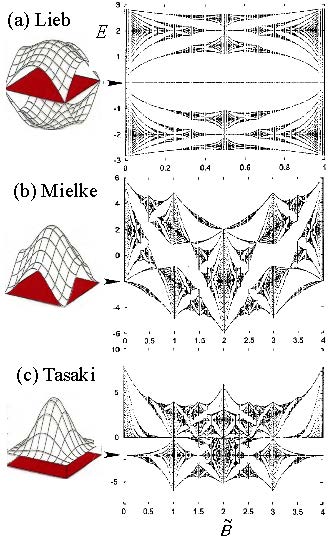}
\caption{
Hofstadter butterflies (energy spectra in an external magnetic field $B$) for Lieb (a), Mielke (b) and Tasaki (c) models [H. Aoki et al, Phys. Rev. B {\bf 54}, R17296 (1996)]. $\tilde{B}$ is the magnetic flux penetrating a unit cell normalised by the flux quantum $\Phi_0 \equiv h/e$.  Arrows mark the positions of the flat bands (in red in left insets) at $B=0$.
}
\label{fig_Hofstadter}     
\end{center}  
\end{figure}

\clearpage

\section{Flatband ferromagnetism}

\subsection{How flatbands favour spin alignment}

Historically, the physics of the flatband, in the present definition with the 
anomalous Wannier states, was initiated when Lieb 
pointed out in 1989 that the Lieb lattices 
should have an itinerant ferromagnetism when there is a repulsive 
interaction between electrons and when the chemical potential 
is set to the flatband energy\cite{Lieb:1989,Lieb04}. 
By `itinerant' is meant that 
the magnetism occurs when the Fermi energy is 
right within an electronic band.   This theorem is quite remarkable 
for the following reasons:  (i) Ferromagnetism is one of 
the oldest subjects in the condensed matter physics.  
In fact, as Edmund Stoner, well-known for the Stoner factor, 
wrote in the introduction to his textbook\cite{stoner} that 
``It is interesting to notice that the two earliest observed 
electromagnetic phenomena --- permanent magnetism and frictional 
electricity --- are among those which have longest defied completely 
satisfactory explanation." and he goes on to 
mention Thales of Miletus 
({\it c}. 630-550 B.C.) as being attracted by the subject.  
Now, the ferromagnetism comes in two flavours.  One is the ferromagnetism in insulators, and the other is the itinerant (or band) ferromagnetism.  While in the former, 
the magnetism has basically to do with the exchange interactions 
between localised spins, in the latter 
the magnetism occurs over the Bloch-state electrons 
in an electronic band.    
In other words, the magnetism arises because 
the electron-electron interaction 
exerts its effect in a manner dependent on the spin states 
of the Bloch electrons.   Thus the insulating magnets are 
easier to understand as an effect of a ferromagnetic 
spin-spin coupling, but the band ferromagnetism is more 
subtle.  In fact there are very few rigorous theoretical examples 
of the band ferromagnetism.  One is Nagaoka's ferromagnetism (1966), 
which will be exlained.  The other is the flatband 
ferromagnetism (1989).  

We can readily realise that a band ferromagnetism is not 
easy to obtain in {\bf Fig.4}
.   
Stoner's theory (1946) uses a mean-field (Hartree-Fock) picture 
to predict that  a ferromagnetic 
ground state is expected if a dimensionless quantity 
$UD(E_F)$ exceeds unity.  Here $U$ is the strength of 
the repulsive electron-electron interaction, which is 
usually taken to be an on-site repulsion in the Hubbard 
model, while $D(E_F)$ is the density of electronic 
states at the Fermi energy.  Intuitively, the criterion comes 
from an obervation that a spin alignment will 
lower the interaction energy since 
the repulsive interaction is hindered by Pauli's exclusion 
principle, which overcomes, for $UD(E_F)$ above a 
critical value, the enhanced kinetic 
energy from the spin imbalance.  However, it has been realised that 
the more we go beyond the mean-field theory towards 
electron correlation physics, the more stringent the 
critical value becomes.

\begin{figure}[ht]
\begin{center}
  \includegraphics[width=1.0\textwidth]{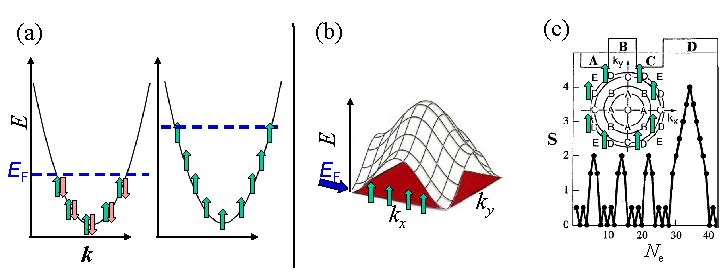}
\caption{
(a) Spin configurations in usual paramagnetic and ferromagnetic 
metals.  Arrows represent electron spins.  (b) Flatband ferromagnetism.  
(c) Generalised Hund's coupling in $k$-space, here exemplified 
for an open-shell Fermi surface in the Hubbard model on a finite square lattice, for which 
the ground-state total spin $S$ is plotted against the number of electrons $N_e$ 
[After K. Kusakabe and H. Aoki, J. Phys. Soc. Jpn {\bf 61}, 1165 (1992)]. 
}
\label{fig_LiebMielkeTasakiHubbard}     
\end{center}  
\end{figure}

Kanamori's theory in the 1960s shows, with the T-matrix 
approximation which becomes valid for dilute electron systems, 
that ferromagnetism does not arise for ordinary lattices 
at least when the band filling is low enough.  
If we increase the filling up to the 
half filling, strong repulsive interactions will 
make the ground state antiferromagnetic at that filling, through the 
kinetic exchange interaction.  
Then Nagoka's theory dictated, again in the 1960s, 
that a ferromagnetic ground state 
emerges if we consider an extreme situation 
in which $U\rightarrow \infty$ limit is taken and 
the 
density of electrons is set at half-filling minus one electron 
(i.e., the doping level from half-filling is infinitesimal)\cite{Nagaoka}.  
The theorem holds rigorously in this situation.  
There is a way to regard the flatband ferromagnetism 
to be related to the Nagaoka ferromagnetism 
as we shall see.

\subsection{Various ways to view the flatband ferromagnetism}

{\bf Lieb's theorem}

Lieb has shown for the Lieb model (repulsive 
Hubbard model on a bipartite lattice where 
the numbers, $N_A, N_B$ respectively, of A and B sublattice sites 
differ from each other) that the ground state at half-filling 
is non-degenerate and has a net magnetisation of 
$S_{\rm tot} = |N_A - N_B|/2$ for $0<U \leq \infty$.  
Here, `non-degenerate' is important, since 
this guarantees that we do not have to worry 
about magnetism being destroyed by some level 
crossing as we change system parameters (here 
the Hubbard $U$).  In the absence of crossing, we can determine the 
magnetisation in the limit of $U\rightarrow \infty$, 
at which the Hubbard model changes into 
the Heisenberg model.  For the Heisenberg model,  
there is Lieb-Mattis theorem\cite{LiebMattis}, 
which asserts that 
the total spin of an antiferromagnetic Heisenberg bipartite model 
should have a net magnetisation of $S_{\rm tot} = |N_A - N_B|/2$ 
(namely, a ferrimagnetic state where the number of 
up spins differs from that of down spins), which is proven 
with  the Perron-Frobenius theorem.  In 
other words, the flatband ferromagnetism 
crosses over to the ferrimagnetic Heisenberg model 
continuously (i.e., without any level crossings among the many-body 
states).  To be precise statistical-mechanically, even in 
the absence of level crossings, we have to examine 
a possibility of a phase transition (a spontaneous 
breaking of symmetry) emerging in 
the thermodynamic limit to infinite systems. 
In the flatband ferromagnetism, we do not have to worry about this, 
since the magnetism is already present in finite 
systems, so the absence of crossings 
implies that the magnetism persists all over finite$\, \leftrightarrow \infty$ 
systems.  

\par
\ \\

{\bf Thouless theory}

While a Mielke model (kagome) has corner-sharing 
triangles, there exists 
an interesting quantum effect for spin physics already for a single triangle 
as noted by 
David Thouless as early as in the 1960s when he discussed 
exchange interactions in solid $^3$He\cite{thouless65}. 
There, he gives a notion of what 
happens when more than one electrons undergo 
cyclic permutations on a parquet such as a triangle. 
As illustrated in {\bf Fig.5},
a cyclic permutation of two electrons on a triangle 
$(i,j,k)$ is expressed as an exchange operation 
$(\bm{\sigma}_i \cdot \bm{\sigma}_j + 
\bm{\sigma}_j \cdot \bm{\sigma}_k + \bm{\sigma}_k \cdot \bm{\sigma}_i)$ 
with $\bm{\sigma}_i$ being the spin operator at site $i$, 
and its coefficient representing the interaction is shown to be 
ferromagnetic.  
\par
\ \\

\begin{figure}[ht]
\begin{center}
  \includegraphics[width=0.6\textwidth]{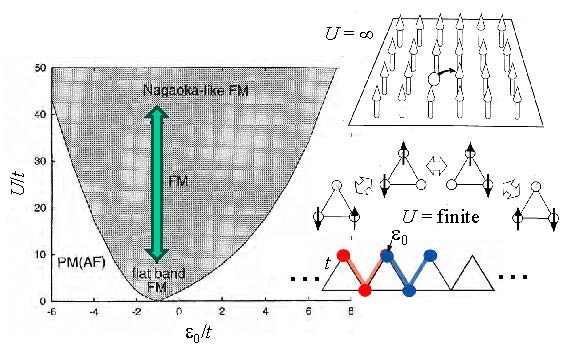}
\caption{
Phase diagram for the triangle 
chain against the Hubbard repulsion $U$ and the level offset $\varepsilon_0$ of the apex site for 1/4 filling (i.e., when the 
flatband, which is realised at $\varepsilon_0=-1$, is half-filled)   [after K. Penc et al, Phys. Rev. B {\bf 54}, 4056 (1996)].  
Insets represent a hopping of a hole for $U\rightarrow \infty$ (top), how a cyclic permutation of electrons on a unit results in a ferromagnetic interaction (middle), and the triangle chain with overlapping Wannier 
orbits (bottom). 
}
\label{fig_FlatVsNagaoka}     
\end{center}  
\end{figure}

{\bf Nagaoka's ferromagnetism meets the flatband ferromagnetism 
via the Perron-Frobenius theorem}

Nagaoka's theorem belongs to precious 
few examples of exact theorems about band ferromagnetism.  
We can then pose a question: would the flatband ferromagnetism 
be related to Nagaoka's in any way?  Curiously, the answer is yes.  
For that, we have to start with the Perron-Frobenius theorem, 
which is a standard topic 
in the undergraduate course on the linear algebra.  
We have stressed that the flatbands are defined as those 
having unorthogonalisable Wannier states.  This 
can be paraphrased as the condition that the 
density matrix ($\rho_{i,j} = \langle\Psi|c^{\dagger}_{i}c_{j}|\Psi \rangle$) is  `indecomposable', 
which means that any two sites $i$ and $j$ 
are connected 
via a series of nonzero matrix elements in the density 
matrix (see {\bf Fig.6}
, inset).  
This condition is also known 
as the `connectivity condition' in the flatband literature.  
Now, the Perron-Frobenius theorem asserts that 
the lowest eigenenergy of an indecomposable 
non-negative (Hamiltonian) matrix corresponds to a single root (Frobenius 
root), where the eigenvector, non-degenerate apart from 
spin degeneracy, has the components that are all nonzero and of 
the same sign.

\begin{figure}[ht]
\begin{center}
  \includegraphics[width=0.6\textwidth]{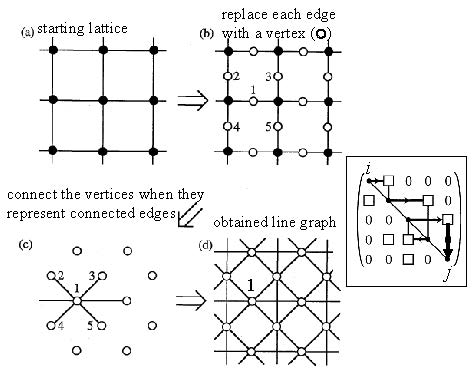}
\caption{
Line-graph construction of flaband lattice models, 
here exemplified  
by starting from a square lattice to obtain 
the checkerboard lattice ($\in$Mielke model).  
Inset shows how we can visualise  
the connectivity condition on the density matrix, where we can go from 
any site $i$ to $j$ via nonzero matrix elements (squares). 
}
\label{fig_linegraphs}     
\end{center}  
\end{figure}

This can be applied to many-body systems such as the Hubbard model 
to explain the flatband ferromagnetism as well as Nagaoka's ferromagnetism.  
In Nagaoka's case, 
the half-filled band (one carrier per site on average) 
is doped with a single hole 
at $U=\infty$.  We can 
show, following an argument by Tasaki\cite{tasaki89}, that motions of the hole generate all 
the possible electron configurations, which realises 
the connectivity condition, and the resulting ground-state 
wavefunction is ferromagnetic and has no nodes.  
In one-body systems, a well-known theorem in quantum mechanics, again in the 
undergraduate level, says that ground-state wavefunctions 
have no nodes (i.e., the whole function has the same sign), and the above theorem for the indecomposable (i.e., flatband and Nagaoka) cases 
is a kind of generalisation to many-body cases, so to speak.

If we want to go away from the Nagaoka limit 
(single hole with $U=\infty$), we can go over to the flatband 
systems, where Lieb's theorem guarantees that 
a half-filled flatband has ferromagnetic ground states 
all over $0<U\leq \infty$ if the lattice satisfies the connectivity 
condition.  
The flatband ferromagnetism treated 
with the Lieb theorem is powerful 
in that, unlike analytic methods such as 
Bethe ansatz, the proof only refers to 
a topological property (connectivity of the density 
matrix), where the physical properties such as magnetism 
can be determined without referring to actual values of the matrix elements, let alone actual values of 
wavefunctions.

Indeed, 
Penc et al\cite{penc96} employed a triangle chain (or sawthooth) (a quasi-1D flatband model 
having overlapping Wannier functions) to obtain a phase diagram 
against $U$ and $\epsilon_0$ (level offset between the bottom  and apex sites), see {\bf Fig.5}
.  There is a wide ferromagnetic region, 
which interpolates between the flatband ferromagnetism and 
Nagaoka's.  They named the large-$U$ regime Nagaoka's, because 
the ferromagnetic 
interaction for two electrons on a triangle we have seen above 
persists on the chain in 
an $1/U$ expansion.  
The crossover between the flatband and Nagaoka's cases 
is in fact reasonable, since in both cases  
Perron-Frobenius theorem guarantees the ferromagnetic ground states.  

\par
\ \\

{\bf The flatband ferromagnetism as generalised Hund's coupling}

We can alternatively view the flatband ferromagnetism as 
a ``generalised Hund's coupling" in the momentum space 
following Kusakabe and Aoki\cite{kusakabeAoki92}, 
see Fig.4
(c).
Usual Hund's theorem is for levels in atoms and molecules, 
and asserts that, when we put electrons on degenerate levels 
that are located at the highest-occupied ones, they 
tend to have aligned spins due to the Hund's exchange interaction 
when we have e.g. two electrons on a doubly-degenerate levels.  
Since these levels do not participate in chemical bonding, 
they are called nonbonding molecular orbitals 
(NBMO) in molecular chemistry.  If we now turn to 
solid-state physics to 
look at the energy levels for a lattice, there are a 
series of degenerate Bloch states that reside on 
concentric circles on the Fermi sea.  
For highest-occupied levels, we can look at 
the Fermi surface, which should have degenerate 
Bloch states in general.  If we calculate the 
ground-state spins in the Hubbard model, for finite systems to discern the 
levels, we can see that a Fermi surface 
tends to have nonzero total spins when 
not fully filled (i.e., an open-shell), at which the 
total spin 
becomes maximal (fully spin-polarised Fermi surface) 
when the surface becomes 
`half-filled', i.e., when the number of electrons on 
the surface is equal to the number of levels there 
(unless the total filling is too close to the half filling).  
In this sense, we can regard the aligned spins as 
coming from Hund's coupling in $k$-space.  
In this manner, we can view the flatband 
ferromagnetism as the generalised Hund's coupling taking place 
on the flatband.

\par
\ \\

{\bf Spin stiffness in the flatband ferromagnetism}

When one deals with ferromagnetism, we have to check 
if the spin stiffness (curvature of the spin-wave dispersion 
in momentum space) is nonzero; otherwise the magnetism 
would vanish when the temperature is raised above $T=0$.  
Indeed, Nagaoka's ferromagnet has an infinitesimal stiffness.  
Since a flatband has a singular dispersion, 
one has also to examine whether the ferromagnetism 
is thermodynamically stable.  We can show that 
the flatband ferromagnetism is stable in both 
the weak-coupling ($U \ll t$) and  
strong-coupling ($U \gg t$) regimes\cite{kusakabeAoki_PRL94}: 
the spin stiffness in flatbands are finite as 
$\sim U$ for weak interactions, and 
$\sim t$ for strong interactions.   This sharply 
contrasts with the spin stiffness 
vanishing like $\sim t^2/U$ for $U\rightarrow \infty$ 
in ordinary lattices, and is 
another effect of the unusual 
Wannier states.  The equation of motion for the 
spin wave can be expressed with  the interaction matrix elements in the 
Bloch basis.  
For the on-site repulsion, the elements are constant 
($U$, the Hubbard interaction) in ordinary models.  
This contrasts with the flatband Lieb, Mielke, and Tasaki models, 
which are multi-band systems, and the matrix elements spanned by 
the flatband Bloch-wavefunctions strongly depend on the momentum 
transfers, which makes the spin stiffness robust.  
In other words, the electrons can `lay-by' 
across the multibands in the electron correlation processes.  

\par
\ \\

{\bf Edge states as flatbands}

There are curious examples of flatbands arising 
from edge states in systems with edges.  
Typically, 
a one-dimensional flatband 
appears in a honeycomb lattice (as in graphene) 
when the sample edge has a zigzag chemical bonds ({\bf Fig.7}
)\cite{fujita}.  
The edge states have a flat dispersion against the wavenumber 
along the zigzag-edge, where the flatband starts from the 
Dirac points in $k$-space.  Since a Dirac point in the Dirac field theory 
has an energy $E=0$, the mode is usually called the zero-mode.   
We can show that the appearance of the flat edge mode is by no means 
an accident.  To show that, we can observe that the existence of the zero-mode is protected\cite{Ryu02} 
by topology, where the topological number is $Z_2$ Berry's phases (sometimes called the Zak phase) rather than the more familiar Chern number.  
A zigzag edge is shown to have a nonzero 
Zak phase of $\pi$,  which gurantees an existence of zero-mode edge states, 
even though the gap closes at Dirac points.  
In showing this theorem, an essential symmetry is the 
chiral symmetry which a honeycomb lattice enjoys.  
We generally call a Hamiltonian $H$ chiral-symmetric when an operator $\gamma$ 
exists with which $H$ 
anticommutes as $\{H, \gamma\}=0$.  Graphene 
posesses this symmetry, which is an outcome of the 
fact that a honeycomb lattice comprises A and B sublattices 
(see Fig.7(a)).  This makes the 
tight-binding Hamiltonian block-offdiagonal with 
blocks labelled by A and B sublattices when the electron 
hopping exist only between nearest neighbours.  
The zigzag edge stands out, since 
the number of A and B sublattice sites differ from each other around 
the edge.  If we look closely at the 
zero-mode, its wavefunction is localised along the edge but 
penetrates exponentially into the bulk.

\begin{figure}[ht]
\begin{center}
  \includegraphics[width=0.94\textwidth]{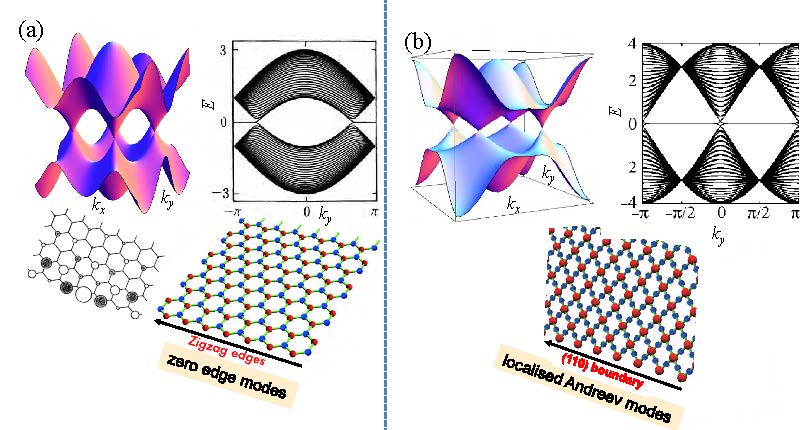}
\caption{
(a) Dispersion (top panels) and the lattice structure with edges (bottom) for graphene [After M. Fujita et al, J. Phys. Soc. Jpn 65, 1920 (1996)]. (b) Dispersion of the Bogoliubov quasiparticle and lattice structure of a 2D d-wave superconductor, represented here as a CuO$_2$ plane [After S. Ryu and Y. Hatsugai, Phys. Rev. Lett. 89, 077002 (2002).]  Dispersion in each case is displayed for the full band structure (left) and a cross section for samples having edges with periodic boundary condition along $y$.   
In (a), the sublattice A(B) is displayed in red (blue), and 
an example of the zero-mode is displayed here for $k_y=\frac{7\pi}{9}$. 
}
\label{fig_edgestates}     
\end{center}  
\end{figure}

An analogous edge mode appears in a superconductor 
with d-wave pairing symmetry where the excitation 
spectrum for the Bogoliubov quasiparticles 
has Dirac cones.  When the sample boundary 
is along (110) crystallographic direction, 
there appears a flatband, which 
arises from Andreev modes (with quantum interference 
between Cooper pairs and electrons/holes) localised 
along the edge\cite{Ryu02}, which also 
has a topological origin and thus robust.\cite{hatsugaiAoki_springer}.  

\clearpage


\subsection{Systematic constructions of flatband models}
\par

{\bf Line-graph and molecular-orbital constructions}

While various lattices belong to the flatband systems, systematic constructions of flatband models are desirable.  
A graphical way is a line-graph construction 
due to Mielke, 
and this is related to the parquet (or molecular-orbital) 
method.  Typical Mielke models have clusters (such as triangles or tetrahedra) 
as building blocks.  Mielke indicated that we can indeed 
have a ``line-graph construction" 
(Fig.6
)\cite{Mielke:1991},  
in which we start from a 
(non-flatband) lattice, with the number of 
vertices (whose number is denoted as $V$) 
connected by bonds ($E$).  Replace each edge with a vertex, 
and connect the vertices with a bond when they originate from 
connected edges.   
The tight-binding model for the 
new lattice is shown to have at least 
$(E-V)$-fold degenerate flatbands.  
The lattices thus generated satisfy the connectivity condition.  
The connected-cluster method can be applied to lattices in arbitrary 
spatial dimensions, such as the pyrochlore lattice.  

Here, it is informative to describe a more general molecular-orbital (MO) 
construction due to Hatsugai and Maruyama\cite{hatsugaiMaruyama11}, 
with silicene as an example.  In this view, a given 
lattice is decomposed into clusters (or MO 
wavefunctions), where any two clusters can share edges 
or vertices.  Then MOs have hopping elements with each other, 
and the secular equation in the tight-binding model 
is shown from the linear algebra to have 
$Z$ flatbands 
with zero eigenenergy with $Z \geq N-M$, where $N$: 
total number of sites, $M$: total number of MOs.  
This is simply shown by counting the rank of the Hamiltonian 
matrix, as in Lieb's argument above.

\begin{figure}[ht]
\begin{center}
  \includegraphics[width=0.8\textwidth]{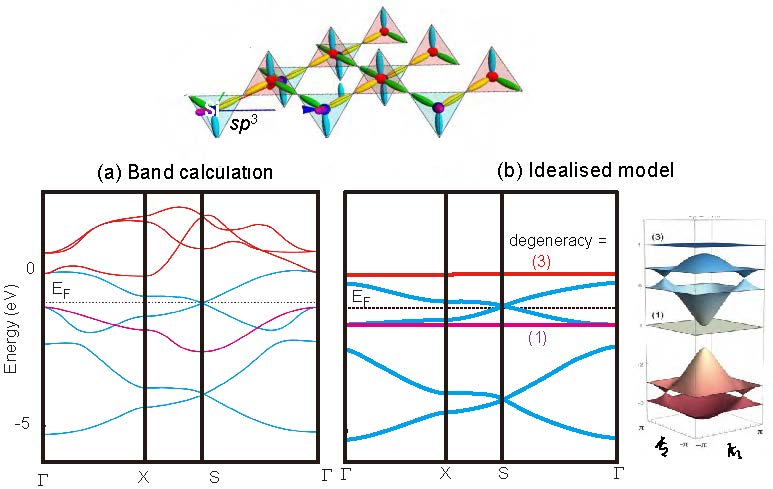}
\caption{
Top inset depicts silicene's structure with each tetrahedron representing $sp^3$ bonds around a Si atom.  (a) Band 
structure of silicene from a first-principles 
calculation, and (b) in an idealised Weaire-Thorpe model as a cross section and a full dispersion.  [After Y. Hatsugai et al, New J. Phys. 17, 025009 (2015).]
}
\label{fig_silicene3}     
\end{center}  
\end{figure}

Hatsugai and coworkers\cite{silicene} applied this to 
silicene ({\bf Fig.8}
), 
which is a silicone version of graphene, by 
first noting that there is a celebrated Weaire-Thorpe model\cite{WeaireThorpe71} 
for amorphous silicone.  
While this model was introduced to study amorphous 
silicone, the band structure in this model 
when Si atoms form a (three-dimensional) crystal 
has a number of flatbands (which are located away 
from $E_F$, hence irrelavant to silicone's 
electronic properties).  Incidentally, it is curious to note that a 
related model was discussed by Dagotto et al in 
the 1980s in discussing lattice fermion systems and Nielsen-Ninomiya 
theorem\cite{Dagotto86}. 
Hatsugai et al applied  (a two-dimensional) Weaire-Thorpe model 
to silicene.  While silicene is a silicone analogue of graphene, 
silicene has a monolayer cut from the 3D silicone 
with a diamond lattice, so that there is 
an important difference from graphene.  Each carbon atom 
in graphene has 3 chemical bonds arising from 
the $sp^2$ hybridisation, leaving one 
$\pi$ orbital relevant to conduction 
(so that the effective model is just a  planar honeycomb 
lattice with each site having one orbital).  
By contrast, silicene comprises basically $sp^3$-bonded tetrahedra, 
so that each site contains four orbitals.  
In other words, silicene has a considerably buckled planar lattice.  
This comes from a quantum chemical difference in silicene from 
graphene, despite silicone sitting just below carbon in the 
periodic table.  Thus silicene as a single layer of 
tetrahedra has a 
Hamiltonian, 
\begin{eqnarray*}
H_{\rm WT}(\bk) =
 \\
\left[ 
\begin{array}{cc}
H_V(0) & V_2 E_4 \\
V_2 E_4 & H_V(\bk)
\end{array}
\right]
\end{eqnarray*} 
where $H_V$ is a $4\times 4$ matrix with a basis $(s,p_x,p_y,p_z)$, 
$E_4$ a unit matrix, and $V_2$ the hopping across the 
adjacent $sp^3$ orbitals.  We can then show that four 
flatbands exist.  
If we compare this with  the 
first-principle band calculation\cite{shiraishi} 
for silicene, they roughly 
resemble each other as a whole, although 
the flatbands are considerably warped in real bands.  

Hatsugai-Maruyama theorem does not require a translational 
symmetry of the lattice, and this is why, in hindsight, Weaire-Thorpe model applies to amorphous Si.  
We can further regard the irrelevance of translational 
invariance to be related to the strange quantum metric (see section on 
that) in flatbands.  

Another comment is: we can alternatively connect the plaquets 
by bonds rather than vertex-sharing, which gives a way to design 
partially-flat bands.  An example is shown 
in {\bf Fig.9}, 
where diamonds are connected by bonds.  In this case, 
partially-flat bands appear due to 
the band structure, where the bands originating from 
the multiple molecular orbitals 
($p_x$-like and $p_y$-like in this example) 
have band repulsion along 
the intersections of $p_x$ band and $p_y$ band due to 
the orbital hybridisation, and this 
gives the partially-flat 
bands.  This model is proposed to have superconductivity with 
an enhanced 
$T_C$\cite{kimuraZenitani02}.  
\par
\ \\

\begin{figure}[ht]
\begin{center}
  \includegraphics[width=0.65\textwidth]{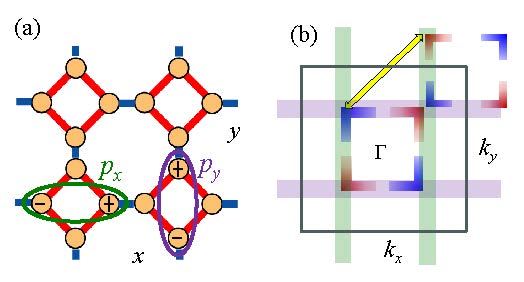}
\caption{
Designed partially-flat band systems is here exemplified by (a) diamonds  connected by bonds.  Each unit (a diamond in this instance) has multiple molecular orbitals 
($p_x$-like and $p_y$-like).  (b) We have band repulsion along the intersections (marked in green and purple) of 
$p_x$ and $p_y$ bands, which results in partially-flat bands.  Squares represent Fermi pockets, a yellow arrow a nesting vector.  
[After T. Kimura et al, Phys. Rev. B 66, 212505 (2002).]  Incidentally, this lattice appears in Ioannis Keppleri: 
{\it Harmonices Mvndi} (1619), who was among the  pioneers 
of crystal structure analysis.
}
\label{fig_ZenitaniLattice}     
\end{center}  
\end{figure}

{\bf Graphene nanomesh construction}

Graphene is interesting in its own right, but 
if we modify the system by introducing a superstructure 
with a long period 
such as a periodic perforation ({\bf Fig.10}
), we can systematically 
control the band structure that encompasses 
flatbands as well as Dirac cones.  This was shown 
by Shima and Aoki as early as in the 1990s\cite{shimaAoki93}.  
The system are later dubbed `graphene nanomeshes', 
and attempts at fabricating the system continue.

\begin{figure}[ht]
\begin{center}
  \includegraphics[width=0.7\textwidth]{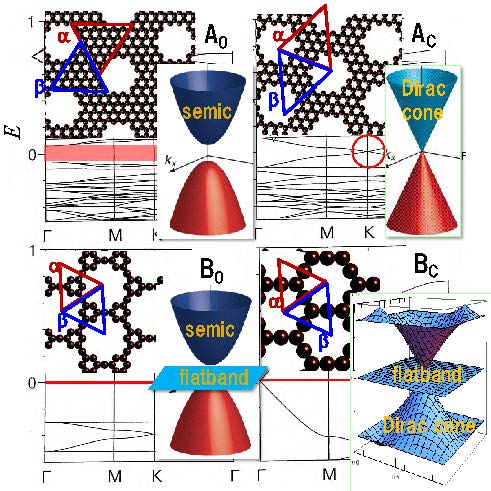}
\caption{
Typical crystal and band structures for the four group-theoretical 
classes [A$_0$: semiconductor, A$_{\rm C}$: Dirac cone, 
B$_0$: semiconductor + flat band, B$_{\rm c}$: Dirac cone 
+ flat band] in the long-period graphene\cite{shimaAoki93}. 
The flatbands are 
marked with horizontal red lines, $\alpha, \beta$ indicate 
the molecular units, 
whose structure determine the class.
}
\label{fig_shimaAoki}     
\end{center}  
\end{figure}

In short, we can classify all the long-period graphene 
with the 2D space group, where, in terms of the band structure, 
there are four classes as \par
\begin{center}
\begin{tabular}{c| c| c| c| c } \hline
 Class &  Formula unit & $\Gamma$ & K & bipartite \\ \hline
  A$_0$ & (C$_{3m})_2$ & & & semiconductor + $n(\geq 0)$ flatband(s) \\ \hline
  A$_{\rm C}$  & (C$_{3m+1})_2$ & & E & semimetal + $n(\geq 0)$ flatband(s) \\ \hline
  B$_0$ & (C$_{3m+3/2})_2$ & A, E & A, E & semiconductor + $n(\geq 3)$ flatbands \\ \hline
  B$_{\rm C}$ & (C$_{3m+5/2})_2$ & A, E & A & semimetal + $n(\geq 1)$ flatband(s) \\ \hline
\end{tabular}
\end{center}
\par
\ \\
Here, the formula unit refers to the carbon structure within the 
unit of the long-period graphene, A (E) means there is a one- 
(two-)dimensional irreducible 
representation in the space group at each of the $\Gamma$ and K 
points in the Brillouin zone, and the band structure 
is indicated for bipartite lattices.  
We can see that flatband(s) {\it have to} exist for classes B$_0$ 
and B$_{\rm C}$, which can be inferred from the number of 
2D reps as combined with the electron-hole symmetry in 
bipartite tight-binding models.   

Incidentally, the Dirac cone dispersion around K point in 
graphene is usually described in terms of a pseudospin-1/2 
SU(2) symmetry (see e.g. Ref.\cite{hatsugaiAoki_springer}).  
In the flatband models, a flatband can intersect the 
Dirac cone as in Fig.10, 
bottom right.  
This might seem a degraded SU(2), but the symmetry is 
preserved, where a change is that the pseudospin is now $S=1$ 
rather than 1/2, which also accounts for the 
triple ($2S+1=3$) degeneracy comprising the flatband and a 
Dirac cone\cite{watanabe11}.

If we go over to three-dimensional systems, Bradlyn et 
al\cite{bradlyn16} have classified the band structure 
of all the 3D space groups to identify 
how the Dirac and Weyl points appear in 3D.  
This is done by looking at 
the maximum degeneracy at the relevant $k$ points 
for each of the 230 3D space groups.  

\par
\ \\

{\bf Flatbands in three dimensions}

\begin{figure}[ht]
\begin{center}
  \includegraphics[width=0.7\textwidth]{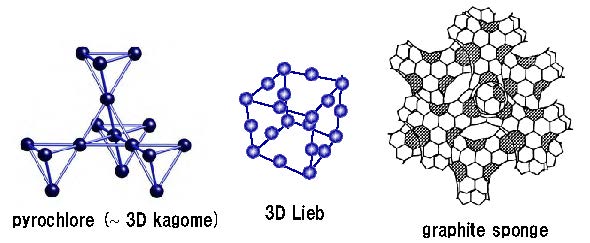}
\caption{
3D flatband models.  [Right panel is from 
M. Fujita et al, Phys. Rev. B {\bf 51}, 13778 (1995)].
}
\label{fig_3Dflatbands}     
\end{center}  
\end{figure}

There are various 
flatband models in three dimensions, as 
displayed in {\bf Fig.11}
.  A typical one is 
pyrochlore lattice (a kind of 3D realisation of kagome) 
belonging to Mielke models.  We can also construct 
3D Lieb and Tasaki models.  The right panel in 
Fig.\ref{fig_3Dflatbands} 
depicts a  ``graphitic sponge" due to 
Fujita et al\cite{graphiticsponge},  
which belongs to a class of zeolite-like structure constructed from graphene sheets, 
and some of the structures accommodate flatbands.  
Units can contain odd-membered rings as far as the Kekul\'{e} rule is 
satisfied\cite{kusakabe97}. 
One way to fabricate the sponges would be zeorite-templated carbons\cite{koretsune}.

\clearpage

\section{Flatband superconductivity}

\subsection{Some basics --- single- vs multi-bands and incipient bands}

Kicked off by the discovery of the high-Tc superconductivity 
in cuprates in 1986, there is a long hisotry of experimental 
and theoretical studies for high Tc superconductivity.  
Concomitantly, there are various attempts at 
searching for materials with higher Tc, 
materials designs, 
and theoretical proposals for mechanisms 
of superconductivity.  
Importantly in the present context, 
there is a recent surge of interests in flatband 
superconductivity, which has now 
become an active field indeed.  
For a review, see \cite{aoki_stripes}.   
A key question of course is: can flatbands favour superconductivity?  
It is becoming increasingly clear that they can, and 
the present section describes this both intuitively 
and analytically.

We have two starting situations: one is 
the attractive electron-electron interactions, and 
the other is the repulsive interactions.   
For the former, T\"{o}rm\"{a}'s group has shown that a flatband
can indeed favour superconductivity when the band 
has non-trivial quantum geometry, 
with the superfluid weight lower-bounded
by the topological number\cite{torma}.  This will be elaborated in 
section `Topological flatbands and quantum-metric implications' below.

For repulsive interactions,
on the other hand, a key question is how the presence of flat bands affects electron correlation processes.  
For flatband superconductivity (SC) with 
repulsive interaction, there are basically 
two essential settings: 

(i) Single-band vs two-band (or multi-band) systems:  
Here, we should not confuse this with 
single-orbital vs multi-orbital systems, 
since, even when we start from a single-{\it orbital} system, 
we can have multi-{\it band} structures if the crystal 
structure is e.g. non-Bravais.  
Multi-orbital considerations are important 
especially when we consider compounds typically 
comprising transition metals, and we shall come back 
to this point in the relevant item in 
section `Candidate materials for flatbands', 
but in the present section we concentrate 
on single-orbital cases.   

\begin{figure}[ht]
\begin{center}
  \includegraphics[width=1.0\textwidth]{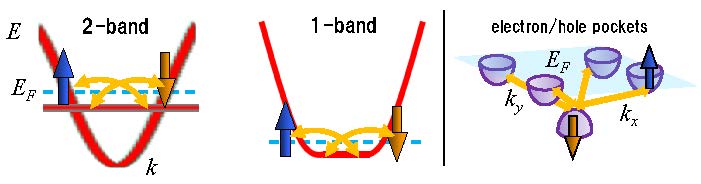}
\caption{
Schematics of the incipient flatband in the 
2-band case (left), and the partially-flat 1-band (middle).  Orange 
arrows stand for pair hopping processes.  
Right panel depicts the incipient pocket for $s_{\pm}$ pairing in FeSe. 
}
\label{fig_incipient2}     
\end{center}  
\end{figure}

(ii) The band structure configurations: 
The analysis of flatband SC has to be done with care, 
because simplistic thoughts are often inadequate.  
For instance, a flatband has a diverging density 
of states, which might seem to give a high Tc in the 
simplest BCS theory, but this does not apply 
because self-energy corrections, which arise 
inevitably in electron correlation, also 
blow up for a large density of states, thereby making  the 
quasiparticles short-lived and degrades SC.  Is there a wayout? 
This is exactly where the notion of the {\it incipient} 
flatband comes in: 
when the flatband is close to, but somewhat away from, 
the Fermi energy $E_F$ (which situation is called incipient), 
the SC is shown to be significantly enhanced.  We can look into this 
for both of the one- and multi-band cases, see {\bf Fig.}\ref{fig_incipient2}.

\begin{figure}[ht]
\begin{center}
  \includegraphics[width=0.55\textwidth]{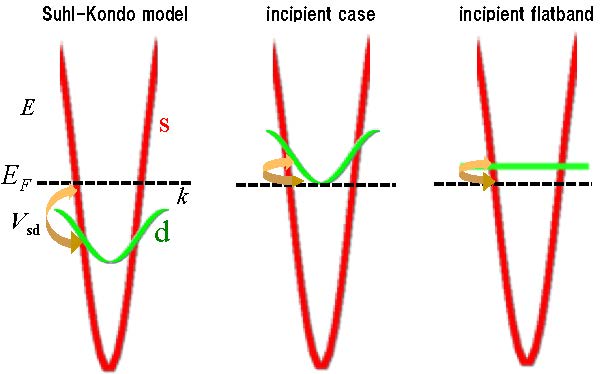}
\caption{
Schematics of the Suhl-Kondo model (left), 
its incipient case (middle), and an incipient flatband (right). 
Orange arrows stand for pair hopping processes.
}
\label{fig_incipient3}     
\end{center}  
\end{figure}

Intuitively, how an incipient band favours SC may be 
first examined in terms of the well-known Suhl-Kondo mechanism\cite{SuhlKondo} 
({\bf Fig.}\ref{fig_incipient3}). 
They have shown, within the BCS formalism, 
that SC occurring on the s band in 
a system comprising s and d bands is enhanced if 
there is an interaction (pair-scattering) $V_{sd}$  between 
the bands.  The increment in Tc is 
\[
\delta T_C \sim V_{sd}^2/|\epsilon_{sd}|
\]
in the leading order in $V_{sd}$, where $\epsilon_{sd}$ 
is the band offset between s and d bands.  So this 
enhancement is always positive (regardless of the 
sign of $V_{sd}$), and exists even when 
the d band is fully-filled or empty.

While the original 
Suhl-Kondo theory assumes that the s band's 
superconductivity comes from attractive interactions 
as in the conventional SCs, the notion 
can be extended to repulsive interactions, where 
an important difference is that we have anisotropic pairings 
in the latter.  
A question is whether a flatband 
can enhance Tc. We shall see for repulisive electron-electron interactions 
that 
higher Tc can indeed arise when the flatband is incipient in the 
two-band case, or 
the flat portion is incipient in the one-band case.  
Enhanced Tc also occurs for 
attractions for incipient flatbands.
The mechanism is from the pair-scattering processes 
between the flat band/portion and the dispersive ones in 
both cases.  
The terminology ``incipient" was often used in the community of the 
iron-based superconductors, 
as in FeSe for the incipient $s_{\pm}$ 
pairing involving the hole band sunk below $E_F$, 
but originally the concept of 
the incipient bands was earlier introduced in a much more general context 
for the cases including flat (or narrow) bands 
by Kuroki et al\cite{narrowwide}.

\subsection{(Flat+dispersive) two-band superconductivity from repulsion}

Let us start with the two-band case, 
where a flatband accompanies a dispersive one.  
If we introduce a repulsive Hubbard interaction on such lattices, 
the basic idea is: even when the Cooper pairs are mainly formed 
on the dispersive band, there exist quantum mechanical 
{\it virtual pair-scattering} processes in which pairs are scattered 
across the dispersive and flat bands ({\bf Fig.13}). 
If we go beyond BCS framework to examine the pair-scattering, 
we can introduce Green's function $G$ 
(which is a matrix for multiband systems), 
and superconductivity is examined with the (linearised) Eliashberg equation, 
\begin{eqnarray}
\lambda \Delta_{l_1l_4}(k)&=&-\frac{T}{N}\sum_q
\sum_{l_2l_3l_5l_6}V_{l_1 l_2 l_3 l_4}(q)\nonumber\\
&\times& G_{l_2l_5}(k-q)\Delta_{l_5l_6}(k-q)
G_{l_3l_6}(q-k).
\end{eqnarray}
Here, $\Delta$ is the gap function matrix spanned by the 
band index $\ell$, 
$k \equiv (\bm{k},i\omega_n)$ with $\omega_n$ being 
the Matsubara frequency, $\lambda$ is the eigenvalue of the 
Eliashberg equation, and the interaction tensor 
$V_{l_1 l_2 l_3 l_4}(q)$ comes from
\begin{equation}
\hat{V}^s(q)=\frac{3}{2}\hat{S}\hat{\chi}_s(q)\hat{S}
-\frac{1}{2}\hat{C}\hat{\chi}_c(q)\hat{C}
\end{equation}
in an abridged expression for the (spin-singlet) pairing interaction, where 
$\hat{S}$ is the spin susceptibility and $\hat{C}$ is the charge 
susceptibility\cite{ikeda}.

\begin{figure}[ht]
\begin{center}
  \includegraphics[width=0.9\textwidth]{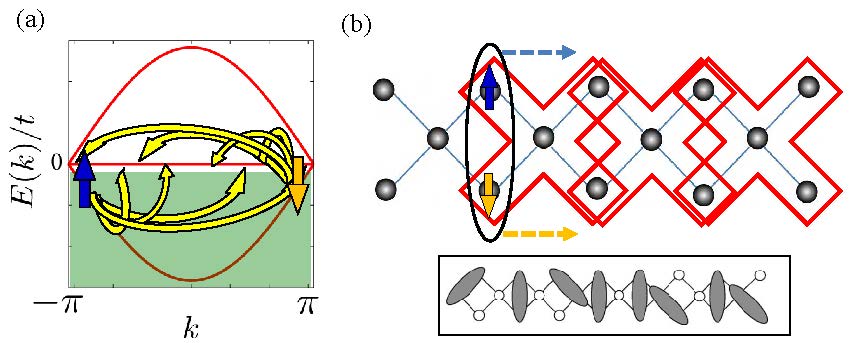}
\caption{
(a) Band structure of the diamond chain, with pair hopping channels represented by yellow arrows.  Green region indicates the occupied band when the flatband is incipient.  (b) Diamond chain, with the unorthogonalisable Wannier functions (red crosses) and a Cooper pair (ellipse) displayed.  [After K. Kobayashi et al, Phys. Rev. B {\bf 94}, 214501 (2016).] Bottom inset is an RVB state with ellipses being spin singlets, from R.R. Montenegro-Filho et al, Phys. Rev. B 74, 125117 (2006).  
}
\label{fig_diamondChain1}     
\end{center}  
\end{figure}

\begin{figure}[h]
\begin{center}
  \includegraphics[width=0.5\textwidth]{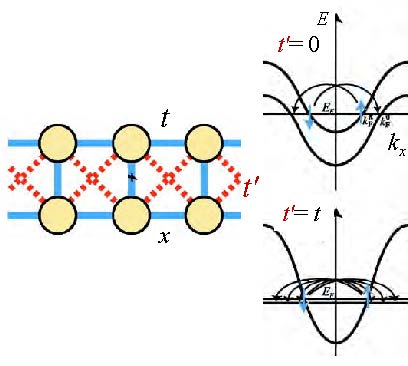}
\caption{A way to realise narrow-wide band systems 
by tuning the hopping $t'$ in a cross-linked ladder (left).  
The width of the 
second band is then varied (right), where arrows represent pair-hoppings. [K. Kuroki et al, Phys. Rev. B {\bf 72}, 212509 (2005)].
}
\label{fig_narrowWide}     
\end{center}  
\end{figure}

As a simplest possible 
quasi-1D flatband model,
Kobayashi et al 
considered the diamond chain, where diamonds are connected 
into a chain ({\bf Fig.}\ref{fig_diamondChain1})\cite{kobayashi}.  
One-body band dispersion consists 
of a flatband sandwitched between two dispersive ones.  
This model is intimately related with 
Kuroki et al's work\cite{narrowwide} cited above, who considered 
the model comprising a 
narrow (or flat in a limiting case) and a wide band 
({\bf Fig.}\ref{fig_narrowWide}).

Since the diamond chain is quasi-1D, 
we can employ DMRG (density-matrix renormalisation group, a method 
capable of treating strong correlations), 
which shows numerically 
that we do have enhanced pairing when the flatband 
is incipient and when the strength of the repulsion $U$ 
is intermediate ($U \simeq 4t$ 
with $t$: nearest-neighbour one-electron hopping).  
We can identify, from the pair correlation function, that the pair is spin-singlet and formed across the apex 
sites of each diamond.  

\begin{figure}[ht]
\begin{center}
  \includegraphics[width=0.6\textwidth]{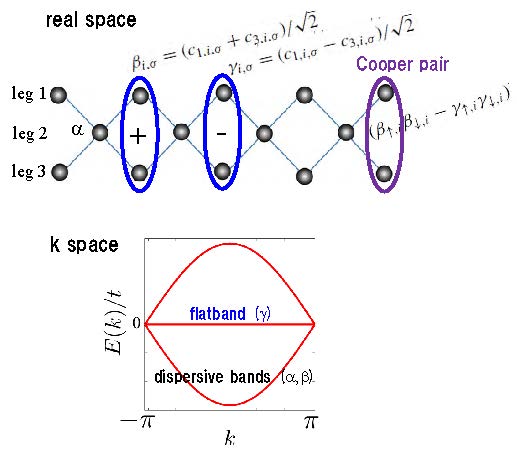}
\caption{
For the diamond chain, the basis functions 
comprise the bonding ($\beta$) and anti-bonding ($\gamma$) states across the top and bottom apexes. The Cooper pair is formed across 
$\beta$ and $\gamma$.  In $k$-space, 
the flatband comes from the $\gamma$ states, while the 
dispersive band from $\beta$ and the middle-leg ($\alpha$). [K. Kobayashi et al, Phys. Rev. B {\bf 94}, 214501 (2016).] 
}
\label{fig_diamondChain2}     
\end{center}  
\end{figure}

For analytic studies, 
we can take a basis as shown in {\bf Fig.}\ref{fig_diamondChain2} 
in terms of the bonding state across the top (leg 1) and bottom (leg 3) 
apex sites [$\beta_{i,\sigma}=(c_{1,i,\sigma}+c_{3,i,\sigma})/\sqrt{2}$] and the antibonding one [$\gamma_{i,\sigma}=(c_{1,i,\sigma}-c_{3,i,\sigma})/\sqrt{2}$].  In $k$-space, 
the flatband comes from the $\gamma$ states, while the 
dispersive band from $\beta$ and the middle-leg ($\alpha$) 
states.  The Cooper pair is expressed as 
\[
(\beta_{\uparrow,i}\beta_{\downarrow,i}-
\gamma_{\uparrow,i}\gamma_{\downarrow,i}).  
\]
If we rewrite the Hamiltonian in this basis, the interaction 
part is shown to contain a pair-scattering term,
\[
\frac{U}{2} \sum_{i} \left( \beta^{\dagger}_{\downarrow,i}
                     \beta^{\dagger}_{\uparrow,i}
\gamma_{\uparrow,i}\gamma_{\downarrow,i} 
+ {\rm H.c.}\right).
\]
This occurs precisely across the 
flat and dispersive bands, whose magnitude is remarkably 
large (half the original interaction $U$).  
If we note the minus sign in the Cooper pair expression, 
the pairing is seen to be $s_{\pm}$-wave between the 
flat and dispersive bands.

\begin{figure}[ht]
\begin{center}
  \includegraphics[width=0.9\textwidth]{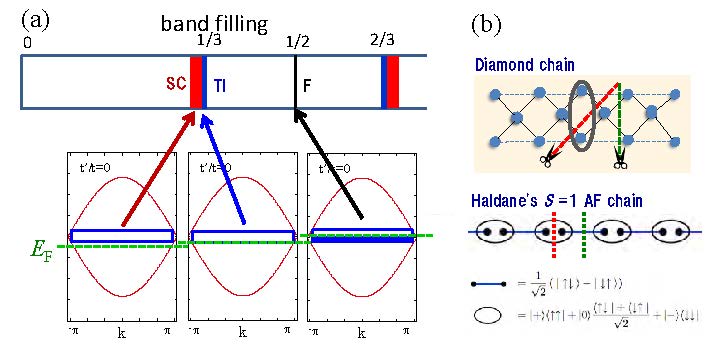}
\caption{
For the diamond chain, (a) phase diagram against band filling, with SC: 
superconductivity, TI: topological insulator, F: ferromagnetism.  Lower inset 
schematically indicates the filling.  
(b) To discern whether the system is topological, we can cut the system differently (red and green lines) to see the resultant 
quantum states, as in Haldane's spin $S=1$ chain.
}
\label{fig_diamondChain3}     
\end{center}  
\end{figure}

We also notice that flatbands accommodate anomalously 
strong quantum {\it entanglements}, which can be deduced 
from the fact that we have to take unusually large number ($\sim 1500$) 
of states in DMRG for convergence in the diamond chain 
even for moderate interaction strengths.  
The large 
entanglements may be reflected in a peculiar 
resonating valence bond (RVB) states where spin-singlet 
pairs are extended over large distances as proposed 
by Montenegro-Filho et al\cite{montenegro}, 
(Fig.\ref{fig_diamondChain1}, inset).  
Large entanglements may be also related to 
the phase diagram against band filling ({\bf Fig.}\ref{fig_diamondChain3}), 
where 
the superconductivity sits right next to a 
topological insulator (TI) that occurs when the dispersive 
band is just fully filled and the flatband is 
just empty.  TI is indeed detected from entanglement 
spectra and also from topological edge states, which is a situation 
very similar to the TI in the celebrated Haldane's 
$S=1$ antiferromagnetic chain.

If we more closely look at the pair correlation function, 
there is a pairing along the chain direction, 
whose correlation function is subdominant but 
has a sign opposite to the dominant one.   
In the ladder physics, 
a pairing correlation that has opposite signs 
between $x$ and $y$ directions is considered to be a precursor 
of a d-wave pairing in two dimensions\cite{noack94}.  
In this sense, the diamond chain has a 
precursory d-wave.  
In ordinary ladder systems, 
the pair correlation function at long distances 
tends to exhibit oscillations, which is related 
with the Fermi-point effect involving the Fermi
wavenumber $k_F$. 
By contrast, the diamond chain is free from this, which should be 
an effect due to the band being  flat.  

\begin{figure}[ht]
\begin{center}
  \includegraphics[width=0.86\textwidth]{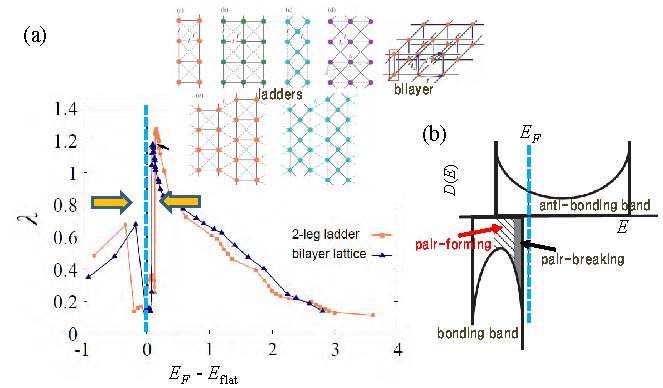}
\caption{
(a) Eliashberg equation's eigenvalue $\lambda$ (a measure of Tc) against the separation between the flatband energy and $E_F$, calculated for various narrow-wide band systems as indicated in the inset.  Orange arrows mark the Tc peak-to-peak 
distance.  (b) Low-energy spin fluctuations (grey area) that tend to degrade SC and higher-energy ones (hatched) that tend to enhance SC are schematically shown on the density of states for the anti-bonding and incipient bonding bands for a ladder.  [after K. Matsumoto et al, J. Phys. Soc. Jpn {\bf 89}, 044709 (2020).]
}
\label{fig_incipient5}     
\end{center}  
\end{figure}

In the phase diagram, SC appears when $E_F$ is slightly below 
the flatband, so this is typical of the incipient SC.  
For the incipient flatband SC in general, 
you can raise a question: can we quantify 
the energy separation between the flatband and $E_F$ 
required to have higher Tc?  
Recently, Kuroki's group has shown the following for 
various flatband models ({\bf Fig.}\ref{fig_incipient5})\cite{matsumoto20}.   
Numerical estimates of Tc, in 
terms of the eigenvalue, $\lambda$, of the Eliashberg equation, 
show a 
general trend for sharply enhanced SC in these models 
as the Fermi 
energy $E_F$ approaches the flatband energy, 
where $\lambda$ (a measure of Tc) 
is considerably larger than in usual cuprate models. 
If $E_F$ is too close to the flatband, however, this causes a 
sharp {\it dip}.  
Details of the width of the dip (i.e., distance between the peaks) 
depends on the Hubbard repulsion $U$, 
the degree of warping of the flatband due to many-body effects, and 
the lattice structure, and these are 
related with the self-energy effect in the flatband 
system.  There, they identify that the key factor 
is the momentum-integrated dynamical spin susceptibility's imaginary part, 
$\sum_{\bm{q}}\, {\rm Im}\, \chi(\bm{q},\omega)$, whose peak as a function of 
$\omega$ gives a measure of the dip width.  
In a wider context, it has been known in the high Tc community 
that low-energy ($\omega \sim < 0.1t$) spin fluctuations act to 
degrade SC, as shown for d- and s-wave SCs, 
while high-energy ($0.1t \sim < \omega \sim < t$) spin fluctuations 
tend to enhance SC\cite{millis88}.  The present case is the flatband version of 
that occurrence, where the flatband helps since the pair-forming 
energy region can be tuned with respect to the incipiently-positioned flatband energy.

Some comments are due: First, the peaks described 
above are intuitively natural and  ingenious, 
since in that situation 
the spin fluctuations having 
finite energies ($\sim$ the energy offset between 
$E_F$ and the flatband) act as a significant 
pairing glue without arousing a strong quasiparticle renormalisation 
that would usually degrade SC.  
Second, while the pair-scattering between the flat and 
dispersive bands sounds a weak-coupling perturbative picture, 
the notion works even in the strong-coupling regime.  
We have already seen that the diamond chain was treated 
with DMRG (a method accommodating strong-coupling).  
Dynamical cluster quantum Monte Carlo (a non-perturbative method) is also used by Maier et al\cite{maier}, 
and variational Monte Carlo by Kuroki's group\cite{kurokiVMC} 
to show the flatband SC. 
Thirdly, diamond chain's band structure can be tuned 
as shown by Vollhardt and coworkers, who have 
applied magnetic fluxes to systematically probe the 
flatband ferromagnetism, and they show that electron itineracy as well as 
magnetism can be controlled by the flux\cite{vollhardt07}.

\begin{figure}[ht]
\begin{center}
  \includegraphics[width=0.54\textwidth]{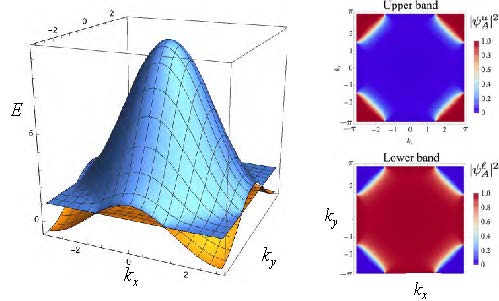}
\caption{
A class of models where a flatband 
intersects a dispersive one, here exemplified for a 
tetragonal case.  Right panel shows the orbital components 
of the upper and lower bands in the lattice comprising 
A and B sublattices.  
[After T. Misumi and H. Aoki, Phys. Rev. B {\bf 96}, 155137 (2017).]
}
\label{fig_Misumi}     
\end{center}  
\end{figure}

A different question is the following.  Usually, the flatband is either 
located at the Dirac point 
(where the density of states vanishes) as in Lieb model, 
touches the bottom of a dispersive band as in Mielke 
model, or separated from the dispersive band by a gap 
as in Tasaki model.  
So a natural question is: can we make a flatband 
located right within a dispersive one, which may favour SC.  
Misumi and Aoki have shown that we can indeed systematically 
extend the flatband models to a class of models 
where 
a flat band pierces a dispersive one 
by tuning distant hoppings in 2D lattices 
as shown in {\bf Fig.}\ref{fig_Misumi}\cite{misumi}.  
The connectivity condition and the unorthogonalisable 
Wannier states are still present.  The orbital components 
can also be tuned by deforming the lattice model to 
promote the pair-scattering between the flat and 
dispersive bands.

\subsection{Partially-flat one-band superconductivity}

\begin{figure}[ht]
\begin{center}
  \includegraphics[width=1.0\textwidth]{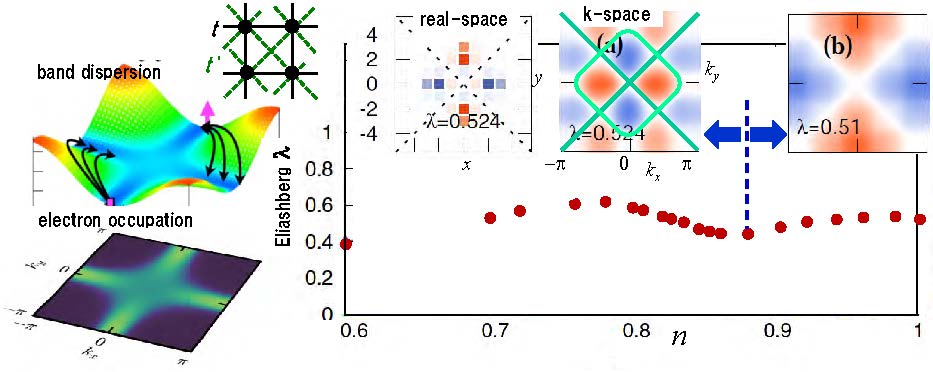}
\caption{
Left: Band dispersion of a frustrated ($t'=-0.5t$) square lattice, with arrows indicating pair hopping processes to and from the flat part of the dispersion, along with the electron occupation when the flat part is close to but below $E_F$.  Right: The eigenvalue $\lambda$ of the Eliashberg equation against band filling $n$ obtained with FLEX+DMFT.  Gap function in k- and real spaces are displayed for each of the double dome.  [After S. Sayyad et al, Phys. Rev. B 101, 014501 (2020).]
}
\label{fig_partiallyflat1}     
\end{center}  
\end{figure}

Let us now turn to a question: 
for the flatband SC, do we have to have two-band systems 
or can single-band systems accommodate a flatband SC as well?  
Sayyad et al have shown that, even in one-band models, 
we have an enhanced SC if the band has a flat region 
in the Brillouin zone, which they call 
a {\it partially-flat} band\cite{sayyad} ({\bf Fig.}\ref{fig_partiallyflat1}).  
For such a model, Huang et al\cite{vaezi} have 
studied superconductivity for attractive interaction 
$U$ and Mott insulation for repulsive $U$ in the 
Hubbard model with the determinantal quantum Monte Carlo 
(DQMC) method. 
%

Let us look into the SC for repulsive interactions.   
We can take  (Fig.\ref{fig_partiallyflat1}) 
a $t$-$t'$ model on a square lattice 
with a second-neighbour hopping $t'\simeq -0.5t$ (which 
may be called maximally-frustrated case) to have 
a wide flat portion along $k_x \simeq 0$ and $k_y \simeq 0$.  
With FLEX+DMFT method (the fluctuation-exchange approximation 
combined with the dynamical 
mean-field theory), 
Sayyad et al have shown that strong 
correlation effects emerge well below half-filling 
and even for small repulsive $U$ unlike in ordinary bands.  
Intuitively, this comes from the electrons 
crammed into the flat portion (Fig.\ref{fig_partiallyflat1}, 
bottom left inset).  
The spin susceptibility, $\chi_S$, is shown to 
exhibit large and wide ridges in $k$-space. 
Concomitantly, superconductivity (as measured by 
the eigenvalue, $\lambda$, 
of the Eliashberg equation) as a function of the 
band filling in Fig.\ref{fig_partiallyflat1} 
exhibits a {\it double-dome} structure for 
the dominant spin-singlet pairing.  
The peak on the smaller-filling 
side represents a gap function that 
has an unusually larger number of nodes in $k$-space than in the usual 
d-wave, while the peak on the larger 
filling side represents a usual d-wave pairing. 
If we look at the pairing in real space, the case 
of large number of nodes is traslated to unusually extended 
pairs in real space.

Another interesting observation is: even in normal 
states, {\it non-Fermi liquid} properties are observed.  
A Fermi liquid would have a self-energy $\Sigma$ that behaves as 
${\rm Im}\;\Sigma(\omega) \sim \omega^2$ (on the real frequency axis) or 
${\rm Im}\;\Sigma(i\omega) \sim i\omega$ (on the Matsubara axis).  
The exponent of $\omega$ in the partially-flat bands 
is shown to be about half the usual 
values\cite{sayyad,werner16}. 
Intuitively, the non-Fermi liquid properties 
may be considered to arise from nonlocal (entangled) 
interactions in flatbands. 
Thus the superconductivity in partially-flat bands 
takes place right in the non-Fermi liquid.  
Non-Fermi liquid properties are also theoretically 
indicated for two-band systems such as the 
Lieb lattice\cite{kumar21}.

\begin{figure}[ht]
\begin{center}
  \includegraphics[width=1.0\textwidth]{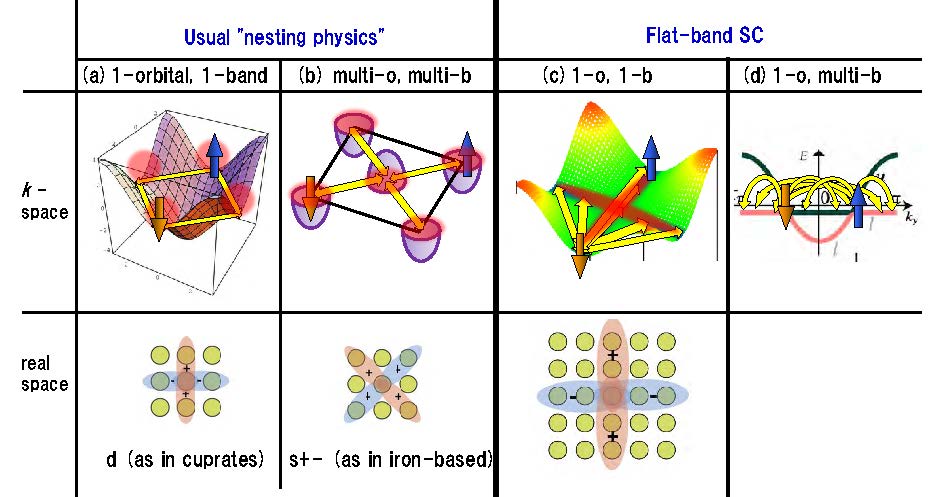}
\caption{
Schematics of (a) the ordinary single-orbital, one-band systems (typically for a d-wave 
SC), (b) multi-orbital, multi-band systems (here for s$_{\pm}$), both with 
specific ``hot spots" (in red) across which the nesting vectors (yellow arrows) designate how pairs (blue and cyan arrows) hop.  
These are contrasted with (c) flat-band systems for 1-orbital, 1-band cases, and 
(d) 1-orbital, multi-band cases. Bottom row displays pairs in real space.
}
\label{fig_flatbSCtable}     
\end{center}  
\end{figure}

Conceptually, these findings are intriguing in that 
the pairing mechanism goes beyond the conventional ``nesting physics". 
This is summarised in {\bf Fig.}\ref{fig_flatbSCtable} 
for electron-mechanism superconductivity from repulsion: Usually, we have well-defined 
nesting vectors, which determine the pairing symmetry. 
The nesting primarily works 
across the ``hot spots", 
which are exemplified by the anti-nodal 
regions in single-orbital, one-band systems as 
in the d-wave SC in the cuprates, or by the 
electron and hole pockets in multi-orbital, 
multi-band systems as in the $s_{\pm}$-wave 
in the iron-based\cite{hosonoKuroki}.   In stark contrast, 
partially-flat bands 
have a {\it bunch} of pair-scattering 
channels, which gives the broadly-peaked 
spin structures and the associated SC.  
This also accounts for the double dome that has 
not very sharp peaks in the partially-flat bands.  
Namely, some interference may arise in pair scatterings 
that exist over a bunch of channels 
from the coexisting dispersive and flat parts within the 
same band.

\begin{figure}[ht]
\begin{center}
  \includegraphics[width=0.2\textwidth]{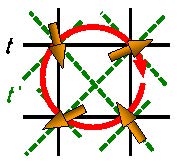}
\caption{
A ring-exchange spin 
interaction is schematically shown on a 
lattice with the diagonal hopping $t' \neq 0$.
}
\label{fig_ringexchange}     
\end{center}  
\end{figure}

As for intuitive elementary mechanism for the peculiar pairing in 
partial flatbands, one possibility is to consider 
``ring-exchange" interactions that work 
for more than three spins ({\bf Fig.}\ref{fig_ringexchange}), 
such as 
\[
\sum_{i,j,k,l}(\bm{S}_i \cdot \bm{S}_j)(\bm{S}_k \cdot \bm{S}_l)
\]
for four spins.  While this class of interactions 
exists for ordinary lattices in higher-order 
expansion in $1/U$ with the coefficient for the above 
expession $\sim t^4/U^3$, introduction of the 
second-neighbour $t'$ produces extra terms\cite{delannoy09}. 
This effect is expected to be stronger 
in the partial flatbands 
for larger $t'$ with intensified frustration.

Another topic related with SC in frustrated lattices is 
that a nematic SC.  Namely, Sayyad et al\cite{sayyad23} 
pointed out that, 
if we consider the triangular lattice, electronic states 
can distort themselves from the many-body 
repulsive interaction, thereby lowering their symmetry 
below that of the lattice.  In general, this kind of 
many-body effect is long known as Pomeranchuk instability, 
and the resulting electronic states are called electron nematicity.  
What Sayyad et al found is that, in the frustrated 
(triangular) lattice (having $C_6$ point-group symmetry), 
nematic electron states emerge with the symmetry lowered to 
$C_2$, and that this enhances SC (almost doubles Tc).  
The resultant pairing symmetry 
is ($d_{x2-y2} + s_{x2+y2} + d_{xy}$) pairing.  
A physical reason for this is that 
the pairing interaction becomes significantly enhanced 
by the nematicity in triangular lattice, 
which contrasts with the square (non-frustrated) lattice where 
the leading (first) order correction from the nematicity 
to the Eliashberg 
equation identically vanishes.

%

\clearpage

\subsection{How the flatbands favour SC}

We can examine several relevant points and open questions on this.  

\begin{figure}[ht]
\begin{center}
  \includegraphics[width=1.0\textwidth]{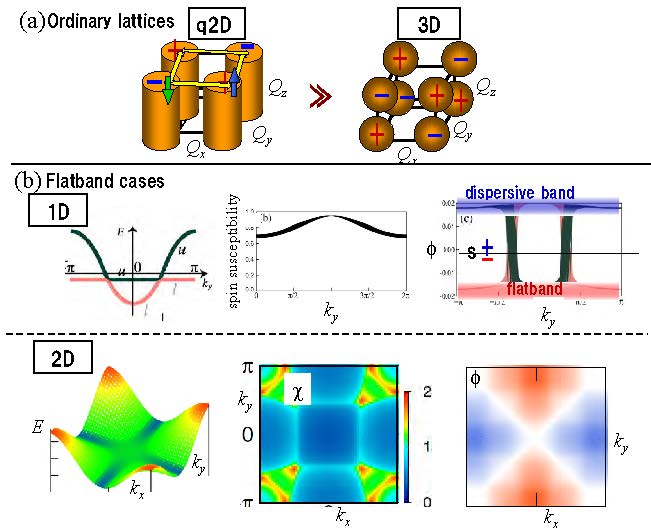}
\caption{
(a) In electron-mechanism superconductivity from repulsion with anisotropic pairs in ordinary lattices, 
the regions in which the spin fluctuation-mediated interaction is large in $k$ space are highlighted in orange 
for layered (quasi-2D) and 3D structures, where ${\Vec Q}$
is the momentum 
transfer.  (b) Band structure (left column), spin susceptibility $\chi_S$ (middle), 
and gap function $\phi$ (right) are shown for a 1D narrow/wide band system (upper row) [after K. Kuroki et al, Phys. Rev. B 72, 212509 (2005)], and for a 2D partially-flat band system (lower) [after S. Sayyad et al, Phys. Rev. B 
{\bf 101}, 014501 (2020)].
}
\label{fig_2D3Dnesting}     
\end{center}  
\end{figure}

(i) {\it Dimensionality}: Usually, in the electron-mechanism SC employing spin-fluctuation mediated pairing, the pairing is dominated by the 
hot spots as mentioned above,  and this implies that 
the pairing interaction is strong 
specifically in compact regions in $k$-space (e.g., Brillouin 
zone corners for the antiferromagnetic spin fluctuations).  
From a phase volume argument as graphically displayed in 
{\bf Fig.}\ref{fig_2D3Dnesting}, 
we can see that quasi-2D (layered) systems have much higher 
volume fraction in $k$-space contributing to the pairing, 
and hence much more favourable 
than in 3D systems\cite{monthouxArita}.  
This is consistent with the experimental fact that most of the 
recently discovered  superconductors have layered structures.  By contrast, 
the flatband systems have much wider momentum regions for large spin 
susceptibility $\chi_S$ than in nesting-dominated cases, 
see Fig.\ref{fig_2D3Dnesting}, lower panels.  
This has a profound effect on the structure of the gap function.  
In a two-band case, the narrow-wide band model\cite{narrowwide} 
for instance 
has an $s_{\pm}$ wave between the flat(+) and 
dispersive($-$) bands, where each band has a relatively 
homogeneous amplitude in $k$-space, coming from 
a homogeneously large $\chi_S$. In a single-band case, 
a 2D partially-flat band system also exhibits 
a spin structure that spreads over the Brillouin zone, 
which gives a gap function whose absolute amplitude also 
spreads over the Brillouin zone\cite{sayyad}. 
If these tendencies continue in 3D, we can expect that 
3D systems can be as good as 2D systems in the flatband SC, 
evading the usual limitation discussed by Monthoux 
et al and by Arita et al\cite{monthouxArita}.

(ii) {\it Vertex corrections}: In general, the size of $T_C$ in SC arising from 
electron-electron repulsion is very ``low" (two orders of magnitude lower than the 
electronic energy), which is identified 
to mainly come from the vertex correction 
in the pair scattering in usual lattices such as square, 
as shown by Kitatani et al 
with the dynamical vertex approximation (D$\Gamma$A)\cite{kitataniDGA}.  
Thus how the vertex correction works in the flat-band 
systems is an interesting future problem.  

\begin{figure}[ht]
\begin{center}
  \includegraphics[width=0.95\textwidth]{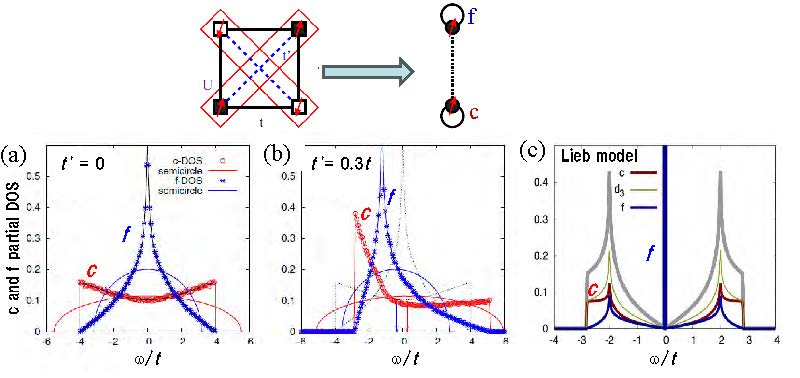}
\caption{
c/f fermion picture is schematically shown, where 
the original system is recast into a Kondo-like system with a 
basis transformation (top left inset), DMFT embedding, and single-site approximation onto a Kondo-like lattice with $c$ and $f$ fermion species 
(top right). Lower panel depicts the $c$ and $f$ partial density of states for (a,b) t-t' square lattice with two values of the 
second-neighbour hopping $t'$ [After P. Werner et al, Phys. Rev. B 94, 245134 (2016)], and for (c) Lieb model 
[After P. Werner and S.A.A. Ghorashi, Phys. Rev. B 111, 045138 (2025)]. 
}
\label{fig_spinfreezing}     
\end{center}  
\end{figure}

(iii)  {\it c/f fermion picture}: 
One way to view the physics would be the c/f fermion transformation 
devised by Werner and coworkers to map 
a single-orbital system onto a kind of Kondo lattice, 
see {\bf Fig.}\ref{fig_spinfreezing}\cite{werner16}.   
There, the original 
system is recast with a basis transformation introducing 
$c$ and $f$ fermion species, followed by 
DMFT embedding and single-site approximation. 
They have examined how the second-neighbour 
$t'$ in a square lattice modifies the situation, 
where the van Hove singularity has an f-character 
as seen in the partial densities of states.  
They have also applied the method to Lieb model\cite{wernerGhorashi}, 
where the flatband mainly supports f character but 
c and f fermions are hybridised due to the 
overlapping Wannier states in flatbands.   
Thus the c/f description sensitively 
reflects the starting band structure, so that it is an interesting 
future work to look into its effect on SC.

\clearpage

\subsection{Repulsive vs attractive models}

{\bf Repulsion-attraction transformation}

While we have so far mainly focussed on repulsive interactions 
and resulting magnetism and superconductivity, 
how about attractive interacrtions?  
In considering this, we can start with noting a curious point which 
connects repulsive and attractive systems in terms of 
magnetism and superconductivity order parameters.  
Namely, for a single-band Hubbard model on a bipartite 
lattice, a repulsive model can be mapped, at 
half-filling, onto 
an attractive model with a unitary transformation\cite{shiba72}. 
The transformation changes order parameters as

\begin{center}
\begin{tabular}{c|l|l}
\hline
            & repulsion & attraction \\ \hline
            & AF(Z)    & CDW \\
Bravais lattices & AF(XY)   & BCS \\
            & F(XY)    & $\eta$-pairing \\ \hline 
            & AF(Z)    & CDW \\
Flatband (Lieb) model   & AF(XY)   & $\eta$-pairing \\
            & F(XY)    & BCS \\ \hline
\end{tabular}
\end{center}
where AF: antiferromagnetism, F: ferromagnetism, CDW: 
charge-density wave, BCS: usual pairing, and $\eta$-pairing: 
a special kind of pairing with nonzero total momentum of a pair.  
From this table, we can see that, first, a ferromagnetism 
in usual lattices with a repulsive interaction 
translates into a superconductivity 
with a strange pairing called $\eta$-pairing for attraction.  
This contrasts with 
flatband models' ferromagnetism 
for a repulsion translating into 
a usual BCS superconductivity for attraction.  
Algebraically, this is known as an extra SU(2) symmetry 
in the Hubbard model\cite{yang89}.  
which translates the spin degeneracy in the ferromagnetism 
into the degeneracy with respect to the 
number of electrons in the BCS state.  
In this sense, the flatband ferromagnetism 
is related in a natural manner to superconductivity when the interaction 
is sign-changed.

\par
\ \\

{\bf Attractively interacting systems with light and heavy masses}

\begin{figure}[ht]
\begin{center}
  \includegraphics[width=0.8\textwidth]{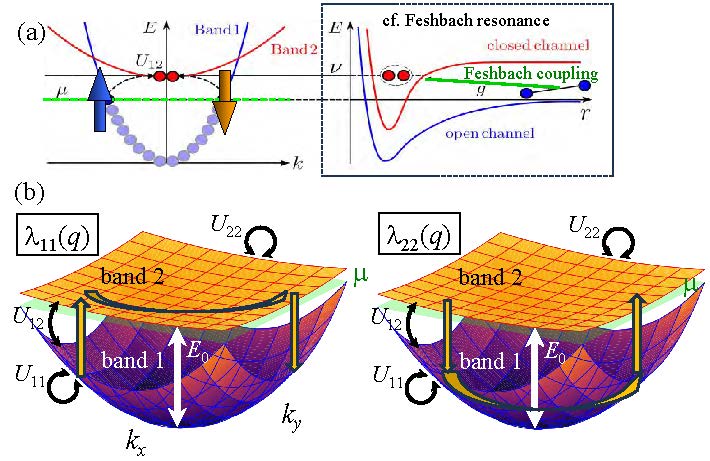}
\caption{
(a) Left: A two-band model with light-mass band 1, and heavy-mass band 2 offset by $E_0$ with an attractive interaction.  Right: Feshbach resonance in atomic physics consisting of open and closed channels, where $\nu$: the energy of the bound state, $g$: Feshbach coupling.   (b) Band 1 has pair-scattering to and from band 2 (left), with band 2 having similar processes (right).  $U_{\alpha \beta}$ is the intra- and inter-band interactions for bands $\alpha,\beta$, and band 2 is incipient, set close to the chemical potential $\mu$.  These result in the Fano-Feshbach-like pairing interaction, whose leading-order expression is 
$\lambda _{\alpha\alpha}(q) = -U_{\alpha\beta}\Pi_{\beta\beta}(q) U_{\beta\alpha}$ with $\Pi$ being the particle-particle correlation function.
}
\label{fig_incipientBCSBEC4}     
\end{center}  
\end{figure}

Now, if we turn to an attractively 
interacting two-band system, we can show that they 
can also harbour an enhanced superconductivity 
when the second band is quasi-flat (with a heavy band mass) and incipient, 
as theoretically shown by Tajima's group with cold-atom systems 
in mind.  They 
traced back its origin to a resonant pair scattering that is 
highlighted by a BCS (Bardeen-Cooper-Schrieffer) to BEC 
(Bose-Einstein condensation) crossover\cite{ochi}.   
By `resonant' is meant the following (see {\bf Fig.}\ref{fig_incipientBCSBEC4}): 
Light-mass band 1 has pair-scattering to and from heavy-mass 
band 2, with band 2 having converse processes, 
and these interband 
pair-exchanges enhance the intraband attraction in each band 
in a Suhl-Kondo mechanism.  There, the pairing 
interaction is shown to specifically intensified 
when the heavy-mass band is  incipient.  
This can be regarded as an (electronic) 
Feshbach resonance, as identified 
from the dependence of the effective interactions and gap functions 
on the position of the chemical potential.  
Feshbach resonance was originally conceived in atomic physics, 
in a situation in which there are 
open and closed channels for atomic scattering, where 
the closed channel is assumed to have a  bound state with an energy $\nu$, 
and the two channels are coupled with a Feshbach coupling  $g$.  
We can then draw an analogy in the electronic resonant pair-scattering, 
where the light-mass band and the heavy-mass (incipient) band correspond,
respectively, to the closed and open channels in atomic physics.  
The Feshbach coupling and the binding energy 
in the closed channel in the latter are translated, respectively, to the interband interaction $U_{12}$ 
and the band offset $E_0$ in the former.  

We can elaborate this in terms of the band-resolved BCS-BEC crossover.  
Cold-atom systems are governed 
by the s-wave scattering length $a$, and, 
as the chemical potential $\mu$ is shifted passing the 
bottom ($E_0$) of the heavy-mass band 2 thereby changing the 
occupation of each band, the light-mass band 1 crosses 
from the BCS regime (with $1/k_0 a^{\rm eff} <0$) 
to a strong-coupling BEC regime ($1/k_0 a^{\rm eff} >0$), 
while band 2 crosses from the unitarity limit ($1/k_0 a^{\rm eff} =0$) to 
weak-coupling BCS regime ($1/k_0 a^{\rm eff} \ll 0$).  
Here $a^{\rm eff}$ defined for each band is the effective scattering length that 
reflects the pair-exchange-induced intraband attraction, 
and $k_{0} \equiv \sqrt{2m_{1}E_{0}}$.

%

When one deals with a BCS-BEC crossover, one has to be careful about how
quantum fluctuations affect the many-body states, i.e., particle-hole 
fluctuations suppressing the pairing in the case of attractive interactions. 
Historically, there is Gor'kov-Melik-Barkhudarov (GMB) formalism for 
treating particle-hole
fluctuations in attractive systems in a continuous space.  
While this was originally devised for one-band (one fermion species) 
systems, Tajima's group extended the formalism to two-band 
systems\cite{tajima24}.  
They find that, while the GMB corrections usually suppress Tc significantly, 
two-band systems with an incipient heavy-mass band can 
have the enhanced pairing, 
which competes with the suppression from the particle-hole
fluctuations.  This results 
in a trade-off leading to a kind of Tc `dome', where 
Tc against the mass ratio $m_2/m_1$ first sharply increases from the 
Feshbach resonance as the ratio is 
increased from unity, then gradually decreases as an effect of particle-hole 
fluctuations. 
When band 2 is incipient, the system plunges into a
strong-coupling regime with the GMB screening vastly suppressed.  
Band 2 can sustain a bound state just below the band bottom depending on 
the relative position of the chemical potential to the band 2 bottom 
as well as on $m_2/m_1$.  The enhanced Tc with suppressed GMB screening 
occurs prominently when the chemical
potential approaches the bound state, and this may be viewed as a Fano-Feshbach
resonance, with its width governed by the pair-exchange interaction.  
Fano resonance is evoked because the band-2 bound state resides 
right in the continuum of band 1.  The relevant Feynman diagrams are shown to 
comprise heavily entangled  particle-particle and particle-hole channels, 
so that the Fano-Feshbach resonance 
dominates both channels, and this may be a rather 
universal feature in multiband superconductivity, especially 
for quasi-flat second bands.

A comment about the band filling, for lattice systems: The incipient narrow/flat band usually refers to full or empty bands near the Fermi level.  Werner and coworkers have demonstrated that\cite{yueWerner22},  even 
when the band is half-filled, doublon-holon fluctuations can 
boost the superconducting Tc for the half-filled attractive bilayer
Hubbard model on the square lattice using dynamical mean-field theory.

\clearpage

\section{Candidate materials for flatbands}

As for candidate materials realising the flatband models, 
various materials have been considered.  
The following is a list of typical ones:

\begin{itemize}
\item A mineral azurite\cite{azurite}.  
This is a famous pigment known from ancient Egyptians and 
Japanese (visit Sanzen-in in Kyoto to admire the pictures 
with this mineral pigment used), see {\bf Fig.}\ref{fig_azurite}.  
Its chemical formula is Cu$_3$(CO$_3)_2$(OH)$_2$ and 
the crystal structure comprises chains, where 
Cu$^{2+}$ ions are coupled via OH$^-$, and 
an effective model is considered to be the diamond chain.  
The material has been 
investiaged as a quantum magnet, since an antiferromagnetic spin system 
of $S=1/2$ on the diamond chain has been theoretically shown to 
accommodate magnetism characteristic of frustrated 
quantum magnet, and one manifestation has been observed as a magnetisation 
plateau against external magnetic field.  
We have to note, however, that there are a number of assumptions and limiting procedures to 
map the material onto the diamond chain, 
so their validity will have to be examined.  
The material is insulating, and it would be interesting 
if we can dope it as in the ladder cuprate.

\begin{figure}[ht]
\begin{center}
  \includegraphics[width=0.6\textwidth]{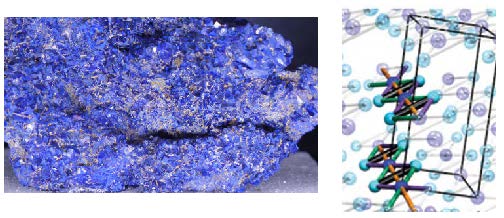}
\caption{
Left: Mineral azurite, photo taken by the present author at the Geological museum, AIST, Tsukuba.  Right: A theoretically optimised crystal structure, with Cu$^{2+}$  ions forming dimers (cyan) and monomers (blue), which gives narrow Cu 3d bands around $E_F$ [H. Jeschke et al, Phys. Rev. Lett. 106, 217201 (2011)].
}
\label{fig_azurite}     
\end{center}  
\end{figure}

\item Organic kagome materials.  There are various inorganic 
kagome materials known, such as  Herbertsmithite which 
is a rhombohedral mineral with a 
chemical formula ZnCu$_3$(OH)$_6$Cl$_2$. 
See the item `Multi-orbital systems' below.  
We can also conceive organic kagome materials.  
One way is to 
consider two-dimensional metal-organic frameworks (MOFs), 
see {\bf Fig.}\ref{fig_MOF}.  
MOFs are usually 3D systems, utilised as catalysts, chemical sieves etc, 
but Yamada et al\cite{yamada16} designed a two-dimensional MOF 
using an organic molecule (phenalenyl) as ligands to 
put heavy-element (Au) atoms into a kagome network.  
For the right choice of the constituents, 
we can obtain a half-filled flatband.  Phenalenyl is 
a radical having  unpaired electrons, which helps to put 
the Fermi energy right at the flatband.  The density functional theory 
indeed shows a ferromagnetic state.  With Au being a heavy element, 
a significant spin-orbit coupling opens a topological gap between 
the flatband and a dispersive one, making the (nearly) flatband topological 
with  a nonzero Chern number.
 So we end up with an organic ferromagnetic and topological flatband.  
Chemists are attempting at fabricating such 2D MOSs\cite{sakamoto22}.

\begin{figure}[ht]
\begin{center}
  \includegraphics[width=0.78\textwidth]{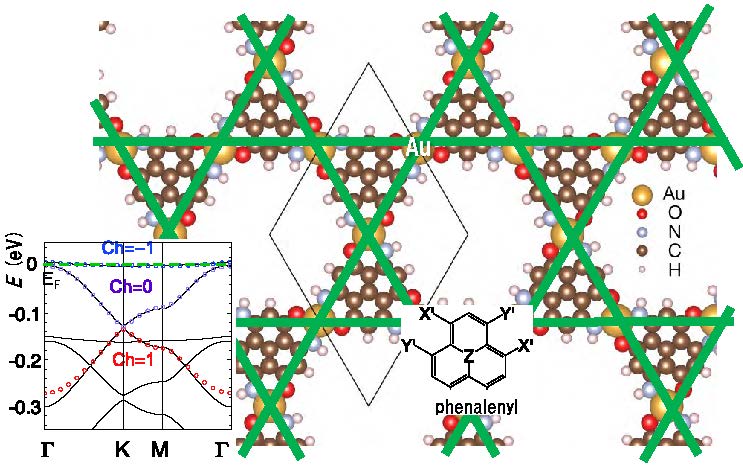}
\caption{
A designed metal-organic framework (MOF).  
Green lines highlight the kagome structure, and an organic 
ligand phenalenyl is displayed.  Inset shows the band structure 
when the spin-orbit interaction is taken into account, with the 
topological Chern numbers displayed.  [After 
M.G. Yamada et al, Phys. Rev. B {\bf 94}, 081102(R) (2016).]
}
\label{fig_MOF}     
\end{center}  
\end{figure}

\item ``Hidden ladders" 
in Ruddlesden-Popper compounds such as 
Sr$_3${\it TM}$_2$O$_7$ ({\it TM}: transition metal such as Mo): 
Ogura et al\cite{ogura17}  have 
theoretically predicted that, 
in a bilayer transition-metal compound in the  Ruddlesden-Popper 
series, two electronic ladders, with one ladder 
composed of $d_{xz}$ orbitals of the transition metal 
and the other from $d_{yz}$, are hidden, see {\bf Fig.}\ref{fig_Sr3Mo2O7}.  
Band structure calculations indeed exhibit flat parts characteristic 
of ladders.  Band-filling dependence of the
eigenvalue $\lambda$ of the Eliashberg equation obtained 
with FLEX shows that there is a Tc dome peaked in the filling region 
where the flatband is incipient.  The 
value of $\lambda$ there is similar to those of a 
cuprate HgBa$_2$CuO$_4 \,(T_C \simeq 90$ K).

\begin{figure}[ht]
\begin{center}
  \includegraphics[width=0.8\textwidth]{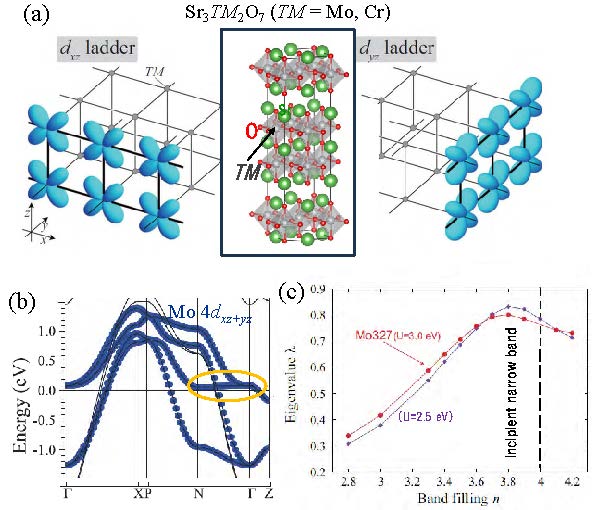}
\caption{
(a) ``Hidden ladders" composed of $d_{xz}$ (left panel) and $d_{yz}$ (right) orbitals in the bilayer Ruddlesden-Popper compounds Sr$_3${\it TM}$_2$O$_7$ ({\it TM}: transition metal). Crystal structure is shown in the centre.  (b) Band structure of Sr$_3$Mo$_2$O$_7$, where the weight of $d_{xz}, d_{yz}$ components are highlighted and the flat part is marked with an ellipse.  (c) Band-filling dependence of the eigenvalue $\lambda$ of the Eliashberg equation for two values of the 
repulsion $U$. The filling around which the flatband is incipient is marked, 
and a vertical dashed line the stoichiometric point.   [After D. Ogura et al, Phys. Rev. B 96, 184513 (2017).]
}
\label{fig_Sr3Mo2O7}     
\end{center}  
\end{figure}

\item Pyrochore Sn and Pb compounds: 
Hase's group\cite{hase} 
proposed that the oxides Sn$_2T_2$O$_7$ ($T$ = Nb,Ta) 
with the pyrochlore structure (Fig.11) 
can be a candidate for the flatband ferromagnetism from 
first-principles band calculations and tight-binding analysis.  
The 
magnetic moments, which arise when hole-doped by N atoms, 
are maily carried by Sn-$s$ and N-$p$ orbitals, 
in constrast to the usual wisdom that $d$ orbitals would be 
required for magnetism.  They have also proposed Pb$_2$Sb$_2$O$_7$, 
where the doping is even unnecessary because of 
a self-doping mechanism that pins the Fermi level at the flatband.

\item Organic conductors:  
Organic solids such as ET-salts come in various crystal 
structures, and, as shown in {\bf Fig.}\ref{fig_tauconductor}, 
one of them 
with a tetragonal structure called $\tau$ -type has a 
peculiar molecular configuration where the ET molecules are 
placed with a face-to-face configuration\cite{tauconductor}.  
If we construct a tight-binding model, this renders 
the diagonal transfer as large as $t' = -t/2$ 
giving a dispersion that has flat portions\cite{arita_tauconductor}.

\begin{figure}[ht]
\begin{center}
  \includegraphics[width=0.5\textwidth]{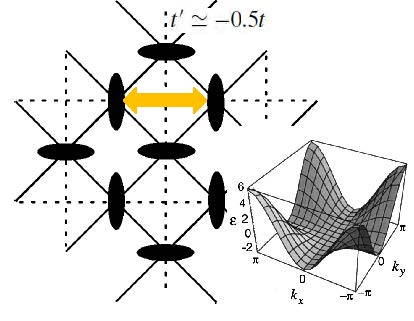}
\caption{
For an ET-salt organic conductor in a tetragonal $\tau$ -type structure, 
a tight-binding model (with each ellipse representing the 
organic molecule in a top view with a face-to-face configuration marked with an arrow, and dashed line the diagonal 
transfer $t'$) with the resultant dispersion shown in inset [After R. Arita et al, Phys. Rev. B 61, 3207 (2000)].     
}
\label{fig_tauconductor}     
\end{center}  
\end{figure}

\item Twisted bilayer graphene: 
Multilayer graphenes are another remarkable arena for 
flatband candidates, where 
partially-flat bands are well recognised in recent years to arise 
particularly in the magic-angle twisted bilayer graphene 
({\bf Fig.}\ref{fig_bilayerGrapheneSC2}), 
for which SC was discovered (along with QHE and Mott insulator)\cite{cao18}. 
Superconductivity $T_C \simeq 1.7$ K is low, but stands 
out in the Uemura plot ($T_C$ against Fermi
temperature for various materials).  
First-principles calculations show 
partially-flat bands\cite{po19}.  
The flatness is suggested to topologically protected against disorders when they 
preserve the chiral symmetry\cite{crepel25}.

\begin{figure}[ht]
\begin{center}
  \includegraphics[width=0.82\textwidth]{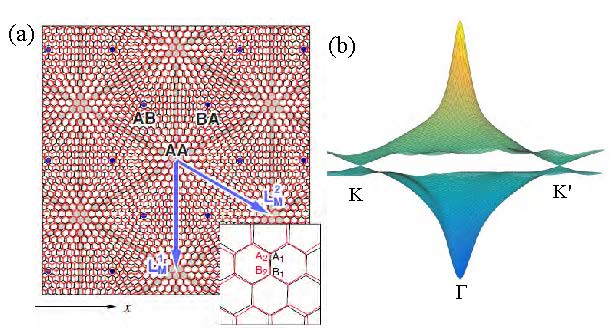}
\caption{
(a) Atomic structure of the twisted bilayer graphene (TBG) with twist angle $\theta (= 3.89$ degrees 
here).   AA, AB and BA stackings are marked, and inset is a blowup.  The primitive lattice vectors of the Moir\'{e} structure are denoted as 
$L_{M}^i \quad (i=1,2)$.  [M. Koshino et al, Phys. Rev. X {\bf 8}, 031087 (2018).]  (b)  An example of theoretical band dispersion 
in $k$-space for a TBG with $\theta = 1.05$ degrees here.  [H.C. Po et al, Phys Rev. B {\bf 99}, 195455 (2019).]
}
\label{fig_bilayerGrapheneSC2}     
\end{center}  
\end{figure}

If we go over to trilayer graphene, 
surface flatbands, localised on the outermost layer, 
are shown to arise in what is called 
ABC-stacking\cite{koshino10}.  
We have to note however that multilayer graphenes  
involve some complications such as a very multi-band 
character coming from the band folding due to the twist.

\item Cold atoms on optical lattices: 
Cold-atom systems where atoms are placed on optical 
lattices are interesting in their own right, 
but also illuminating as emulators of solid-state 
systems.  There are extensive studies to emulate 
such systems as the Hubbard model on a square lattice.  
Needless to say, the cold-atom systems have controllability 
that is much wider than in the solid-state systems, e.g., 
we can tune, with the Feshbach resonance, the interactions not only for their strength but even the sign.  Also, advances in techniques 
for producing various optical lattices have enabled the 
studies towards realising flatband lattices such as 
Lieb and kagome.  One technical hurdle is how to 
lower the temperature below the expected phase 
transition temperature, but technical advances 
are being extended towards that as well.  
See e.g. Ref.\cite{schafer20}. 

\item Multi-orbital systems: 
An important comment on the materials search is that 
most materials are multi-orbital systems, 
with anisotropic orbitals such as d-orbitals in transition metals.  
This means that, even when a 
lattice structure belongs to flatband models, 
the resulting electronic structure should deviate in general 
from the single-orbital ones.  
Conversely, even when the starting 
lattice does not belong to flatband models, 
the resulting electronic structure can have flatbands.  
So let us elaborate on these here.  

{\it p-orbital systems}: 
For $p_x$ and $p_y$ orbitals on 2D lattices, 
one can note how the selection rules for the hopping integrals 
for these orbits affect the band structure\cite{chen23}.

{\it d-orbitals on flatband lattices}: 
There exist kagome compounds such as CsCr$_3$Sb$_5$, which 
belongs to the $A$V$_3$Sb$_5$ ($A$ = K, Rb, Cs) family 
and becomes a superconductor in high pressure.  
There is a theoretical discussion for 
spin fluctuations and superconductivity, on RPA 
level\cite{wu25}.  In this material, $d_{xz}, d_{yz}, d_{x2-y2}$ orbitals are 
relevant around $E_F$, so that, although the Cr atoms form a kagome lattice, 
we have very different band structures, but a DFT 
electronic strucutre indicates incipient partially-flat bands (at $p=5$ GPa).   
A Fermi surface comprising pockets, cylinders and sheets 
having varied orbital characters, and the authors suggest an $s_{\pm}$-wave 
pairing from RPA where an exchange interaction $J$ is taken into account. 
An interesting point is that the authors propose that 
a sublattice-momentum coupling as a driving mechanism for 
spin fluctuations.  Namely, the kagome lattice has 3 sublattices, 
and, if we decompose the sublattice character on the band 
dispersion (including the flatband), patches of different 
characters can be noted, so that spin susceptibility and pairing vertices 
have to be analysed for that. 

There is also an observation of a destructive interference-induced band flattening of partially filled Ni 3d states in a kagome nickelate Ni$_3$In 
for which non-Fermi liquid etc are suggested\cite{ye24}. Here too, anisotropic 
orbtials (mainly $d_{xz}, d_{yz}$) considerebly modify the hopping 
integrals.

Another example is the Lieb-lattice-like cuprate.  
Namely, Li et al found a high Tc cuprate in Ba$_2$CuO$_{3+\delta}$ 
which has heavily (40\%) O-deficient Cu-O plane, but sitll 
has $T_C=73$ K\cite{li19}.  Yamazaki et al studied this material 
theoretically, and they propose an in-plane crystal structure 
that has copper atoms on 
a Lieb lattice\cite{yamazaki20}.  Relevant 
orbitals are $d_{z2}, d_{x2-y2}$, so that flatbands 
do not arise, but the theoretical estimate indicates 
a high Tc from different reasons.

{\it d-orbitals on non-flatband lattices}: 
We have already mentioned in Fig.\ref{fig_Sr3Mo2O7} that an apparently 
non-flatband lattice (bilayer Ruddlesden-Popper compounds) 
can have partially-flat bands from the 
hidden electronic ladders arising from $d_{xz}, d_{yz}$ orbitals.  
Another, earlier example is a generation of kagome from a 
triangular lattice ({\bf Fig.}\ref{fig_koshibae})\cite{koshibae03}.  
Hexagonal cobaltates, such as Na$_x$CoO$_2$ for which superconductivity was 
observed when water-doped, have 
CoO layers consisting of CoO$_6$ octahedra and having Co atoms on a 
triangular lattice.  
From the symmetry, hopping exists only between such adjacent d-orbitals as marked with yellow arrows in the figure, and the resulting tight-binding model is a kagome. The band structure then has a flatband.

\begin{figure}[ht]
\begin{center}
  \includegraphics[width=0.66\textwidth]{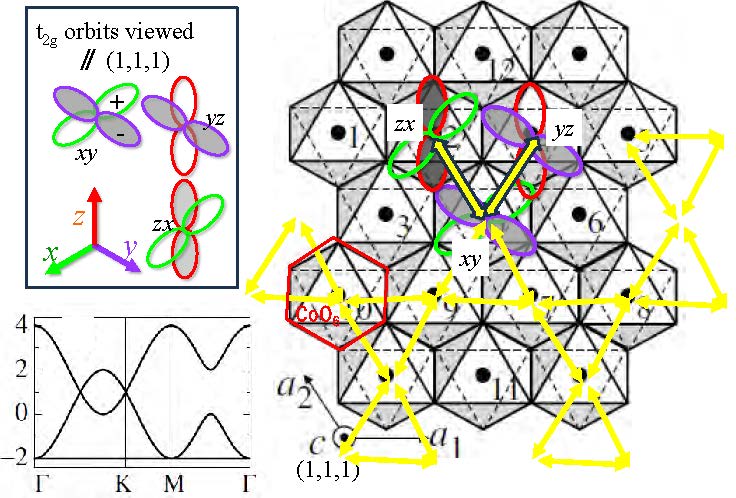}
\caption{
Shown is a CoO layer in a hexagonal cobaltate viewed looking down (1,1,1) direction, where each octahedron represents a CoO$_6$ cluster with a Co atom (black dot) at the centre.  Seen along (1,1,1), three $t_{2g}$ d-orbitals of Co look as in the top left inset. From symmetry, hopping exists only between such adjacent d-orbitals as marked with yellow arrows, and the resulting tight-binding model is a kagome (yellow), although we started from a triangular Co lattice. There are three equivalent kagomes thus generated.  Bottom left inset depicts the band structure.  
[After W. Koshibae and S. Maekawa, Phys. Rev. Lett.  91,  257003 (2003).] 
}
\label{fig_koshibae}     
\end{center}  
\end{figure}

\item Materials search: 
In recent years, papers that intend to comprehensively 
scan and classify flatband materials are beginning to appear.  
See, e.g., 

N. Regnault et al: Catalogue of flat-band stoichiometric
materials\cite{regnault22}, where the authors 
searched for flatbands in two- and three-dimensional stoichiometric materials 
utilising the Inorganic Crystal Structure Database to identify 
in particular materials hosting line-graph or bipartite lattices.

P.M. Neves et al: Crystal net catalog of model flat band 
materials\cite{neves24},   
where the authors develop a high-throughput materials search for flat bands in candidate materials in search for previously unknown motifs.

\end{itemize}

All in all, there is an abundance of flatband possibilities in various crystal structures 
and space groups, and the flatbands may be fairly ubiquitous.  

\clearpage

\section{Topological flatbands and quantum-metric implications}

Physics of topological quantum states has now become one of 
the major fields in condensed-matter physics\cite{aoki_SES2}.    
In the flatband physics, too, topological flatbands 
form a specific class of systems.  
In general, a band is defined as topological 
if the band has a nonzero 
topopological (Chern) number.  
Likewise, a flatband is called topological 
if the flatband has nonzero 
topopological number.  
While topological systems generally have remarkable 
properties, topological flatbands have particular 
interests, since T\"{o}rm\"{a}'s group 
has shown that topological flatbands accommodate 
an outstanding superconductivity (or superfluidity 
if we talk about neutral cold atoms on optical 
lattices)\cite{torma}.  
Namely, they have shown that the 
superfluid weight is ``topologically-protected" as 
\[
D_s \geq (|U|/h^2)|C|
\]
in topological flatbands.  
Here $D_s$ is the superfluid weight (as in 
the optical conductivity $\sigma_1(\omega) = D_s \delta(\omega) + \cdots$), 
$U$ is an attractive Hubbard interaction, 
and $C$ stands for the topological Chern number of the flatband.  
$D_s$ can be explicitly calculated as 
\[
[D_s]_{ij} = \frac{1}{\hbar^2 V}\frac{d^2 \Omega}{dq_i dq_j} |_{{\Vec q}=0},
\]
where $D_s$ is generally a tensor in crystals 
with $i,j$ labelling Cartesian coordinates, $\Omega$ is 
the grand potential in the grand canonical ensemble,
${\Vec q}$ is the wavenumber of the superfluid fluctuation, 
and 
$V$ is the  volume of the system.  
The superfluid weight $D_{\rm s}$ and the superfluid density $n_{\rm s}$ 
are related as $D_{\rm s} = e^2n_{\rm s}/m$, and the supercurrent 
in an external vector potential ${\Vec A}$ is expressed as $\langle j_i \rangle = -[D_s]_{ij}A_j$.

\begin{figure}[ht]
\begin{center}
  \includegraphics[width=0.3\textwidth]{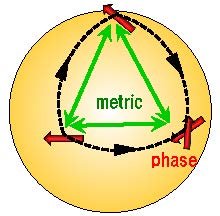}
\caption{
Geometry of wavefunctions is represented for the phase 
(red arrows) and its change against parallel shift (black), along with the 
distance (green), on a space here symbolised by a sphere.  
}
\label{fig_metric}     
\end{center}  
\end{figure}

This can be formulated in terms of the {\it quantum geometric tensor}.  
Namely, the topology of quantum states has recently been 
analysed in terms of the `quantum-metric' description ({\bf Fig.}\ref{fig_metric}).  
The topology does not concern individual wavefunctions, but 
the overall structure (i.e., for the wavefunctions over the entire Brillouin zone in a crystal rather than  the individual Bloch states). 
We can then introduce the quantum geometric tensor (whether or not the band is topological) defined as
\[
{\cal B}_{ij}(\bk) = \langle \partial_i u_{\bk}|\partial_j u_{\bk}\rangle - \langle \partial_i u_{\bk}|u_{\bk}\rangle \langle u_{\bk}|\partial_j u_{\bk}\rangle,
\]
where $\partial_i \equiv \partial/(\partial k_i)$ and $u_{\bk}$ is the Bloch wavefunction with momentum $\bk$. The imaginary part of this quantity, Im$\;{\cal B}_{ij}(\bk) = i\nabla_{\bk} \times \langle u_{\bk}|\nabla_{\bk} u_{\bk}\rangle$, corresponds to Berry's phase, whose integral gives the 
Chern number. On the other hand, the real part, Re$\;{\cal B}_{ij}(\bk) \equiv g_{ij}$, is the quantum metric, and gives a kind of distance between wavefunctions.  This is not just a more precise formulation, but it has been recognised in recent years that the formula is indeed related to observable quantities, and it is being discussed for topological states such as the fractional Chern 
insulators  as well as the  flatband superconductivity. This theme is 
now actively pursued. See P. T\"{o}rm\"{a}'s Ref.\cite{tormaPRL23}.

\begin{figure}[ht]
\begin{center}
  \includegraphics[width=0.8\textwidth]{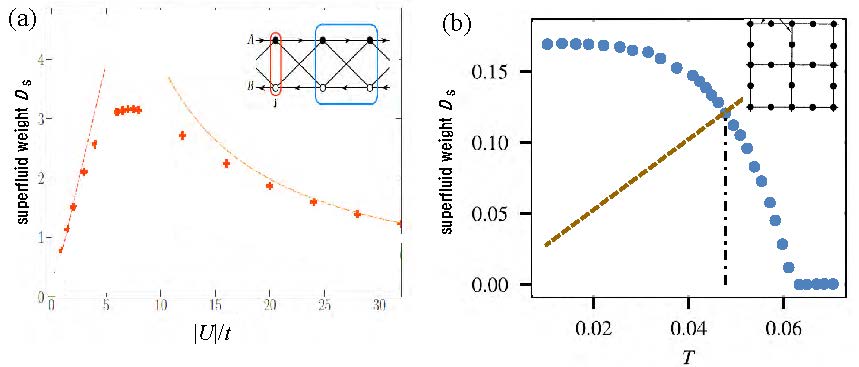}
\caption{
(a) DMRG + ED result for the superfluid weight $D_s$ in the attractive Hubbard model on the Creutz lattice (inset, red: u.c., blue: Wannier state) for the half-filled flatband.  Dashed (chain) line represents $|U| \rightarrow 0 (\infty)$ asymptote. [After R. Mondaini et al, Phys. Rev. B {\bf 98}, 155142 (2018)]  
(b) DMFT result for the superfluid weight $\sqrt{{\rm det}\,D_s}$ in the attractive Hubbard model on the Lieb lattice (inset) for the half-filled flatband.  Dashed line represents $8T/\pi$ for identifying the BKT transition temperature (chain line). [After R.P.S. Penttil\"{a} et al, Comm. Phys. 8, 1 (2025).] 
}
\label{fig_Torma7}     
\end{center}  
\end{figure}

Initially, T\"{o}rm\"{a}'s group calculated 
the superfluid weight in mean-field 
approximations\cite{torma,hutinen22},  
but this was followed by works beyond the mean field.  
Dynamical mean-field theory (DMFT) and exact diagonalisation (ED) 
are used for the attractive Hubbard model on Lieb lattice to show 
that the superfulid weight has a dominant contribution from 
the geometric origin\cite{julku16}.  
Note that, while 
topology guarantees a non-trivial quantum metric, hence 
the topological superconductivity, topology is a sufficient 
(rather than necessary) condition for a topological SC, 
where an example of systems with non-trivial 
quantum metric with zero Chern number is the Lieb lattice.  
A DMFT+ED work for Creutz lattice shows that 
the superfluid weight $D_s$ grows linearly with the attractive interaction 
$|U|$ for small interactions (quite unlike the BCS behaviour of 
$T_C \propto e^{-1/(D(E_F)|U|)}$), with a broad peak against $|U|$ ({\bf Fig.} 
\ref{fig_Torma7}) \cite{mondaini18}. 
It has also been reported that the Pauli (Chandrasekhar-Clogston) 
limit can be violated in flatbands\cite{ghanbari22}. 
These were followed by a quantum Monte Carlo (QMC) result for 
a kagome-like model which confirms $D_s \propto |U|$\cite{huber21}. 
More recently, a DMFT result for the superfluid weight $\sqrt{{\rm det}
\,D_s}$ in 
the attractive Hubbard model on the two-dimensional Lieb lattice 
is used to identify the BKT transition 
temperature, which indicates $T^{\rm BKT} \simeq 0.05t$\cite{penttila25}. 
Note that, first, for 2D systems we have to deal with the Beresinskii-Kosterlitz-Thouless (BKT) 
transition, and, second, in multiband systems, we have to look at $\sqrt{{\rm det}\,
D_s}$ for the superfluid weight tensor (whose expression with 
explicit band indices is given in Ref.\cite{hutinen22}).  
Note that, in multi-band systems, the quantum geometric contributions to $D_s$ 
modifies the relation of $D_s$ with the superfluid density 
from the single-band expression.

As a future problem, while the above studies are for attractive 
interactions, it is a very interesting open question 
to ask how about repulsive Hubbard interactions.
Another theoretical point is that there is a kind of no-go theorem, which 
states that exactly-flat bands cannot be made 
topological if the range of the hopping is finite\cite{chen14}, 
which means that 
we have to turn to the systems having non-trivial quantum geometry 
or else introduce spin-orbit interactions 
within finite-range models.  
Thirdly, there are some attempts at searching for materials with topological flatbands, where e.g. scanning of a first-principles materials database 
is used to identify the compounds that have  topological flatbands 
around the Fermi energy, aided by line-graph flatband models\cite{liu21}.   
The work found a number of candidate two-dimensional flatband materials 
that can become topological when a spin-orbit coupling is introduced, 
where the 
lattice structure is basically kagome and triangle lattices, but 
includes a diamond-octagon lattice.

\clearpage

\section{Flatbands in non-equilibrium}

\subsection{Floquet theory}

This section describes a theoretical proposal for 
realising topological superconductors, in view that 
the frustrated $t$-$t'$ models with partially-flat 
bands may be relevant.  
The proposal starts from non-equilibrium physics, 
i.e., {\it Floquet engineering}, which is 
recognised recently as a route to obtain 
novel quantum states entirely different from 
materials design.  So let us start with an 
introduction for Floquet physics.
 
\begin{figure}[ht]
\begin{center}
  \includegraphics[width=0.8\textwidth]{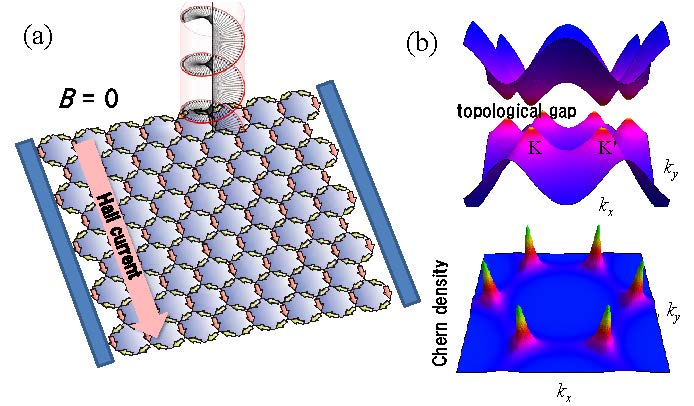}
\caption{
(a) Floquet topological insulator, which arises when graphene is illuminated by a circularly-polarised laser, is schematically shown.  DC Hall current is generated, despite the absence of external magnetic fields.  (b) Topological gap opens dynamically around the Dirac points (upper panel), with topological Chern density emerging there (lower). 
[After T. Oka and H. Aoki, Phys. Rev. B {\bf 79}, 081406(R) (2009).] 
}
\label{fig_Floquet2}     
\end{center}  
\end{figure}

A typical and important way to put a quantum system in non-equilibrium 
is to shine a laser light, which has an oscillating electric field.  
The principle for the Floquet engineering is based on Floquet's theorem for time-periodic modulations 
as conceived by Gaston Floquet in 1883, which precedes 
1928 theorem by Bloch for spatially-periodic 
modulations almost by 
half a century.  
A prime example of Floquet physics is the 
``Floquet topological insulator" proposed by Takashi 
Oka and the present author in 2009\cite{Oka2009}.  
By applying a circularly-polarised light (CPL) to 
honeycomb systems as exemplified by 
graphene ({\bf Fig.}\ref{fig_Floquet2}), we can turn the 
system into a topological state in a dynamical manner.  
The emerging state, having a topological gap, 
is called Floquet topological insulator (FTI).

\begin{figure}[ht]
\begin{center}
  \includegraphics[width=1.0\textwidth]{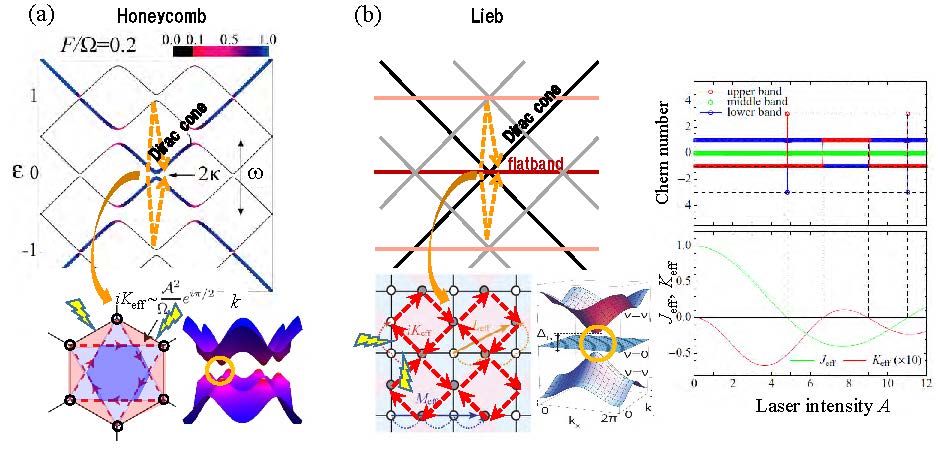}
\caption{
(a) When graphene is illuminated by a circularly-polarised laser, a series of Floquet subbands 
separated by the laser frequency $\omega$ are generated from the original Dirac cone in the energy spectrum, 
here shown against $k$ (measured from each Dirac point). Band repulsion occurs at every band crossing due 
to the Floquet processes, especially at the Dirac point, giving the topological gap (bottom right inset).  From the second-order 
processes (double orange arrows) between the original band and 
the one-photon dressed bands emerges the effective Floquet Hamiltonian that 
exactly coincides with the Haldane's model for the anomalous quantum Hall effect, as shown in the bottom left inset, where   
dashed lines represent second-neighbour (due to the 2nd order Floquet processes as symbolised by yellow lightnings)  hopping 
that is imaginary ($iK_{\rm eff}$ with a positive phase along the arrow, 
negative in the opposite direction).   [After T. Oka and H. Aoki, Phys. Rev. B {\bf 79}, 081406(R) (2009).]  
(b) Corresponding plot for the Lieb lattice.  Avoided crossings are not displayed here, but topological gaps open 
between the flat band and Dirac cone as shown in the bottom inset.  Right panel shows the Chern 
number against laser intensity $A$ for each of three bands in the Lieb model, which is related to the behaviour of 
$iK_{\rm eff}$.  [After T. Mikami et al, Phys. Rev. B {\bf 93}, 144307 (2016).]
}
\label{fig_Floquet6}     
\end{center}  
\end{figure}

As displayed in {\bf Fig.}\ref{fig_Floquet6}(a), the 
FTI is a matter-light combined state, where each 
electron is converted into a superposition 
of the original electron, one-photon dressed one, 
two-photon dressed one, 
..., that are represented by a series of 
replicas (called Floquet 
sidebands) of the original band 
 arising in the Floquet physics, for a 
review, see \cite{OkaKitamura19}.  
The FTI with a topological gap 
exhibits a DC Hall effect despite the modulation being 
AC.  After the theoretical finding, 
Kitagawa et al\cite{Kitagawa2011} have 
pointed out that this result is understandable 
as the effective model for graphene in CPL 
being precisely the celebrated anomalous quantum 
Hall effect (i.e., quantum Hall effect in zero 
magnetic field) originally proposed for the static case 
by Duncan Haldane back 
in 1988\cite{Haldane1988}.  Kitagawa et al have 
shown this in the Floquet formalism 
in the leading (second) order in $1/\omega$ with $\omega$ being 
the frequency of the laser.  
The FTI was then detected in various systems such as 
a surface of a topological insulator and cold-atom systems, 
and in 2019, just a decade after the theoretical prediction, 
McIver and coworkers\cite{McIver2019} 
experimentally detected the Floquet topological 
insulator in graphene itself for which the original 
theoretical proposal was made.

After the experimental report of the FTI in graphene, 
there has been a lot of discussions whether short 
relaxation times in the nonequilibrium dynamics would 
mar Floquet realisations.  Recently, new exerimental 
reports confirmed,  by detecting 
Floquet sidebands, that Floquet physics is indeed 
realised despite ultrafast relaxation\cite{choi25}. 
Technically, we have to be careful in theoretically performing the $1/\omega$ 
expansion to obtain the effective Hamiltonian, since the usually 
employed Floquet-Magnus expansion and van Vleck degenerate perturbation theory 
for AC modulations 
can be ambiguous in systematic higher-order expansions.  Instead, 
we can use the Brillouin-Wigner perturbation theory, which 
gives the whole
infinite series expansion in a consistent and transparent 
manner\cite{Mikami2016}.  

\subsection{Floquet states for flatband systems}

Now, we can apply the Floquet formalism to flatband systems.  
This produces topological gap between the flatband 
and dispersive band(s), and a speciality 
of involvement 
of flatbands gives a wildly-behaving topological Chern numbers.  
Let us first look at Lieb model illuminated by CPL in 
Fig.\ref{fig_Floquet6}(b)\cite{Mikami2016}. 
The Floquet processes are 
at work, this time between the flatband and Dirac cone 
and their Floquet replicas.  The second-neighbour complex hopping, 
$iK_{\rm eff}$ in the leading (second) order 
opens a topological gap above and below the 
flatband, where the flatband remains flat with Lieb model 
being electron-hole symmetric.  If we compute the topological Chern number $C$, 
which describes the anomalous quantum 
Hall effect and is defined for each of the 
three bands in Lieb model, their respective values 
against the laser intensity $A$ behave in a strange manner.  
This comes from 
the behaviour of $iK_{\rm eff}$ and also from the 
Floquet-modified hopping $J_{\rm eff}$ which are respectively 
oscillating functions of $A$.  

\begin{figure}[ht]
\begin{center}
  \includegraphics[width=0.9\textwidth]{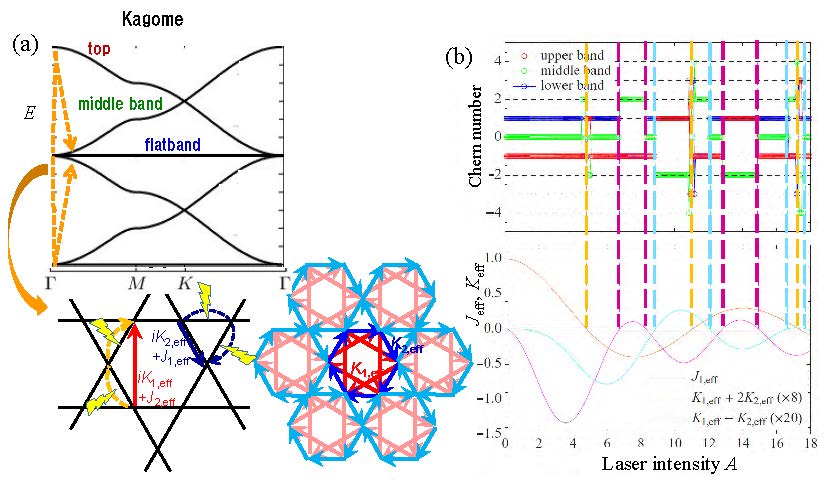}
\caption{
(a) A plot similar to the previous figure for the kagome lattice.  Due to its structure, 
the second-order processes (double orange arrows and double yellow lightnings) produce two kinds of 
Floquet-generated hoppings, one being second-neighbour ($iK_{1,{\rm eff}}$ in red) and the other 
nearest-neighbour  ($iK_{2,{\rm eff}}$ in blue) as described in the bottom insets.   
(b) The Chern number against laser intensity $A$ for each of three bands in the kagome lattice, which comes from the 
behaviour of $K_{1,{\rm eff}}, K_{2,{\rm eff}}$.  [After T. Mikami et al, Phys. Rev. B {\bf 93}, 144307 (2016).]
}
\label{fig_Floquet7}     
\end{center}  
\end{figure}

If we go over to the kagome lattice illuminated by CPL in {\bf Fig.}\ref{fig_Floquet7}, 
the Chern numbers $C_1, C_2, C_3$ for the three bands 
change even more wildly against the laser 
intensity, not only for their absolute magnitudes but 
signs.  An essential
difference in the kagome (an electron-hole {\it a}symmetric model) 
from the Lieb model is that even the flat band (with some warping of the flatness)  has 
nontrivial Chern numbers.

\clearpage

\subsection{Floquet topological superconductivity}

As we have just seen, nonequilibrium, especially the 
Floquet engineering with lasers, has become one of the key pursuits in the condensed-matter physics in looking for or designing new quantum phases. 
While conventional materials 
design, typically for superconductors, tailors the crystal structures and consituent 
elements as combined with carrier doping, pressure, etc, 
the ``non-equilibrium design" should be an entirely different avenue, 
which opens an in-situ 
way to convert the system into new states 
that would be unthinkable in equilibrium.  
So why not utilise this especially for many-body physics such as 
superconductivity to fathom new possibilities.  
Here we describe a theoretical proposal to make an ordinary 
unconventional superconductor into a Floquet topological superconductor.  
Here, `unconventional' means SC with anisotropic pairing 
as in high-Tc cuprates, and `topological' means SC having 
a nozero topological number with broken time-reversal symmetry.  
We shall then discuss an implication for flatband superconductivity.  

\begin{figure}[ht]
\begin{center}
  \includegraphics[width=1.0\textwidth]{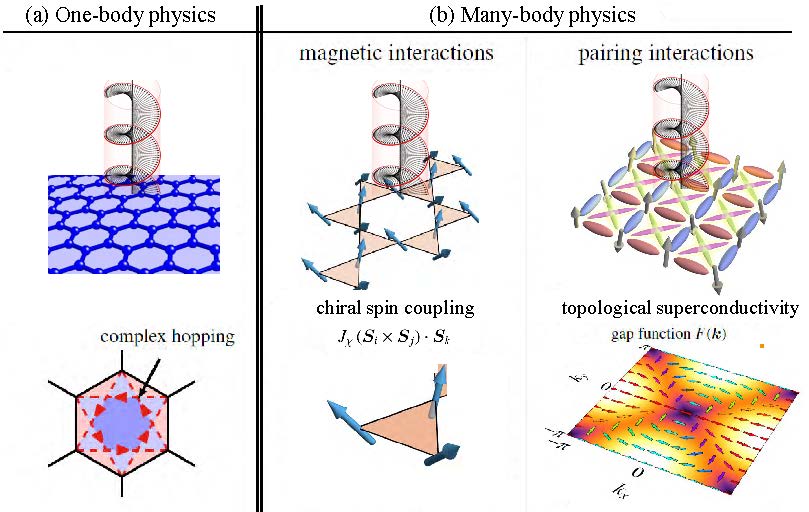}
\caption{
Various Floquet states that emerge when we illuminate circularly-polarised light (CPL) 
to various systems.  (a) One-body physics, where a prime example is CPL-illuminated Dirac fermions as in graphene, 
which induces the Floquet topological insulator.  Hamiltonian is effectively converted, to a Haldane model with  
photon-assisted complex hopping in this case. 
(b) In many-body physics, we can convert magnetic interactions in strongly-correlated 
systems by illuminating CPL, here exemplified by a chiral spin coupling $J_\chi (\bm{S}_i\times\bm{S}_j)\cdot\bm{S}_k$ 
with $\bm{S}_i$ being the spin at site $i$.  For superconductors with strong repulsive interactions, CPL can induce a pairing interaction that is complex and has a different pairing symmetry from the starting 
system, resulting in a topological $(d+id)$-wave superconductivity, as in the bottom right panel.  [After S. Kitamura and H. Aoki, Commun. Phys. {\bf 5}, 174 (2022).]
}
\label{fig_FloquetTopSC2}     
\end{center}  
\end{figure}

Let us start with the {\it many-body Floquet} physics, which now 
encompasses 
a range of quantum phases as listed in {\bf Fig.}\ref{fig_FloquetTopSC2}.  
An essential difference between the one-body Floquet physics and 
many-body Floquet physics is, while a one-body Hamiltonian 
is modified along with the associated band structure, 
laser illumination can change the interaction in a many-body 
Hamiltonian, specifically in strongly-correlated systems.   
If we start from a Mott insulator in that regime for instance, 
a circularly-polarised 
laser induces chiral spin interactions, 
$(\bm{S}_{i}\times \bm{S}_{j})\cdot \bm{S}_{k}$ 
involving three spins,  
for the repulsion $U$ much greater than the electron hopping $t$ 
(Fig.\ref{fig_FloquetTopSC2}(b))\cite{Takayoshi14}.

%

\begin{figure}[ht]
\begin{center}
  \includegraphics[width=0.9\textwidth]{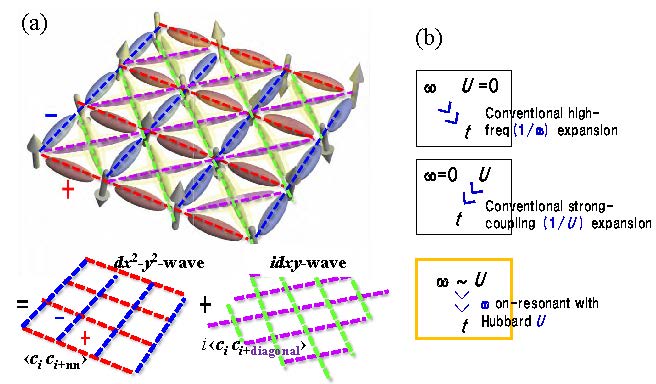}
\caption{
(a) CPL illuminated on a $d_{x^2-y^2}$-wave superconductor produces pairing amplitudes $\langle c_{i\uparrow} c_{j\downarrow}\rangle$ across nearest neighbours (red: positive; blue: negative) along with imaginary diagonal amplitudes (magenta and green), leading to an emergent complexified gap function in $k$ space, 
hence a photo-induced topological $(d_{x^2-y^2}+id_{xy})$ superconductivity. [After S. Kitamura and H. Aoki, Commun. Phys. 
{\bf 5}, 174 (2022).]  (b) In Floquet engineering for the Hubbard model, there are various energy scales ($\omega$: frequency of the laser, $U$: Hubbard repulsion, $t$: electron hopping energy).  Interesting is the case where $\omega \sim U \gg t$.  
[After S. Kitamura et al, Phys. Rev. B {\bf 96}, 014406 (2017).]
}
\label{fig_FloquetTopSC5}     
\end{center}  
\end{figure}

An interesting and rather general observation for Floquet physics 
for many-body systems is that 
a Hubbard model illuminated by laser has three energy 
scales: Hubbard $U$, laser frequency $\omega$, and 
the hopping ($\sim$ bandwidth) $t$, see {\bf Fig.}\ref{fig_FloquetTopSC5}(b). 
Interesting phenomena tend to occur when $\omega \sim U \,(\gg t)$, 
which may be called $\omega$ `on-resonant' with $U$.  
The Floquet-induced chiral spin coupling indeed becomes significant 
when the frequency $\omega$ of 
the laser is close to the Hubbard $U$, exemplifying 
``$U$-$\omega$ resonance", which can be quantified in Floquet equations 
involving $(U-{\rm integer}\times \omega)$ in the energy denominator, 
and is reflected in vastly different behaviours 
between the cases for red- and blue-detuned  $\omega$ 
from $U$.

\begin{figure}[ht]
\begin{center}
  \includegraphics[width=0.7\textwidth]{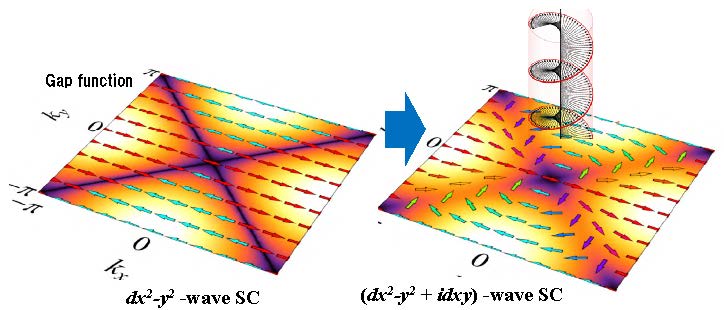}
\caption{
Change of a $d_{x^2-y^2}$-wave SC into a 
$(d_{x^2-y^2}+id_{xy})$-wave SC after an illumination of  circularly-polarised laser, where arrows stand for the phase of the complex gap function.
}
\label{fig_FloquetTopSC3}     
\end{center}  
\end{figure}

Now, for superconductivity, 
Kitamura and Aoki\cite{kitamuraAoki22} 
have shown that an illumination of a circularly-polarised laser 
can change a $d$-wave superconductor to a topological superconductor, 
namely, a ``Floquet 
topological superconductivity" arises, see {\bf Fig.}\ref{fig_FloquetTopSC3}.  
There have been various attempts at realising Floquet topological superconducting states, but an obstacle there is that 
 pairing symmetry 
is hard to be Floquet-controlled in a direct manner, since the gap function does not couple to electromagnetic fields.  In this sense, a Cooper pair is electrically 
neutral.  What Kitamura and Aoki proposed is that we can overcome this by 
exploiting the laser-induced 
{\it interactions} (here the pairing interaction) 
that arise in strong-correlation regime $(U\gg t)$.  
Namely, an illumination of a circularly-polarised light (CPL) 
to the repulsive Hubbard model in the strong-coupling regime 
modifies the pairing interaction, which results in superconductivity 
changed from the usual $d$-wave into a topological $(d+id)$-wave 
(Fig.\ref{fig_FloquetTopSC2}(b)).  The key interaction is 
the two-step correlated hopping 
caused by the CPL, along with the chiral spin coupling caused by the CPL.  The former is dominant, and turns out to remain significant 
even for relatively low frequencies and moderate intensities of the CPL.  
Obtained phase diagram against the laser intensity 
and temperature shows a `Tc dome' against the laser field intensity.

\begin{figure}[ht]
\begin{center}
  \includegraphics[width=0.8\textwidth]{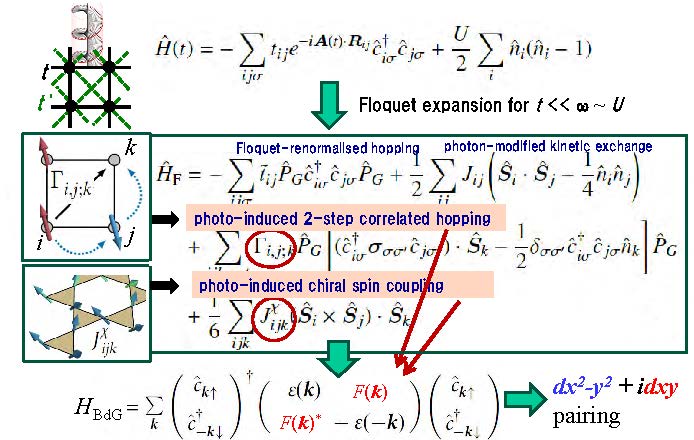}
\caption{
Starting from the Hubbard Hamiltonian on e.g. square lattice in a circularly-polarised laser, a Floquet expansion can be done for $\omega \sim U  \gg t$.  This results in an effective Hamiltonian, which comprises the Floquet-renormalised hopping, photon-modified kinetic exchange interaction, photo-induced two-step correlated hopping $\Gamma$, and photo-induced chiral spin coupling $J^{\chi}$. $\hat{P}_G$ is the Gutzwiller projection. We can then plug these into the Bogoliubov-de Gennes Hamiltonian, and this gives a $(d_{x^2-y^2}+id_{xy})$ pairing.  
[After S. Kitamura and H. Aoki, Commun. Phys. {\bf 5}, 174 (2022).]
}
\label{fig_FloquetTopSC6}     
\end{center}  
\end{figure}

We are not going into technical details here, but the reasoning is as follows: 
If we look at the Bogoliubov-de Gennes equation 
describing SC in {\bf Fig.}\ref{fig_FloquetTopSC6}, 
the off-diagonal terms related to the gap function 
in the Nambu representation 
do not contain the vector potential from the 
laser's electric field (while the diagonal normal terms do). 
This is related a Cooper pair coming from electron and 
hole branches in the BCS picture, and 
this is why we cannot readily have Floquet SC.  
The difficulty can be overcome by evoking 
photon-induced interactions in strong-correlation 
regime\cite{kitamuraAoki22}.  
For strong $U$, while the leading (second) order term in 
$1/U$ expansion is known to give an effective Hamiltonian as 
t-J model (in the static case), we can go to higher orders, where 
the Floquet ($1/\omega$) 
expansion can be performed at the same time.  If we do this 
in the regime $U \sim \omega \gg t$, 
the result is as displayed in  Fig.\ref{fig_FloquetTopSC6}, 
where the higher-order terms consist of 
(i) the photo-induced {\it two-step correlated hopping} $\Gamma$, 
which involves three sites and occurs in the 
influence of the strong repulsion $U$ as well as Pauli's 
exclusion (such as an up-spin at site $i$ hops to 
site $j$, where the electron experiences $U$ with 
down-spin at $j$, which subsequently hops to a third 
site $k$), along with (ii) the photo-induced {\it chiral spin coupling} 
$J_{\chi}$  
which also involves three electrons as mentioned above.  
These do affect the off-diagonal terms in the BdG Hamiltonian, 
and, importantly for CPL, the terms are imaginary (i.e., 
breaks the time reversal).  Thus, when we 
start from an ordinary $d_{x^2-y^2}$-wave SC as in high-Tc 
cuprates, illumination of CPL causes a pairing 
interaction that makes $id_{xy}$ pairing to emerge, 
and we end up with a topological $(d_{x^2-y^2}+id_{xy})$ wave, 
see Fig.\ref{fig_FloquetTopSC5}(a).  
This can occur in ordinary lattices such as square lattice 
(which, within one-body physics, does not support the Floquet topological 
insulator).

The amplitudes of the photo-induced two-step correlated hopping $\Gamma$ 
and the photo-induced chiral spin coupling $J_{\chi}$ are 
respectively very sensitive functions of the circularly-polarised-light amplitude $E$ and driving frequency $\omega$.  This is because 
they involve Bessel's functions of $E/a\omega$ times the 
$U$-$\omega$ resonance factors, with a leading order of $E^4$ 
when Taylor-expanded in $E$.  This may sound fairly complicated, 
but we can note that $|\Gamma|$ is peaked around $E \sim 2\omega/a$
 ($a$: lattice constant) and blows up for $\omega \rightarrow 0$, and that 
$|J_{\chi}|$ blows up for $\omega \rightarrow U$.  
This makes the required laser intensity $E$ 
moderate, despite the relevant process being of the 4th order.  
For an optimal $E$, the $id_{xy}$ component is as large as 
$i\times0.3t\,{\rm sin}(k_x) {\rm sin}(k_y)$, which is comparable with 
the $d_{x2-y2}$ component of $\simeq 0.7t[{\rm cos}(k_x) -{\rm cos}(k_y)]$.

A phase diagram against the laser field intensity E and the 
temperature exhibits a significant region for the CPL-induced 
superconductivity. 
In this calculation we set $\omega$ 
slightly red-detuned from a $U$-$\omega$ resonance, 
for which the two-step correlated hopping $\Gamma$ 
blows up along with $J_{\chi}$.   


Thus we have a way to obtain a topological SC.  
For topological systems in general, we have 
nowadays a well-known classification scheme\cite{classificationTable}. 
There are altogether ten universality classes for topological quantum 
states, and superconducting states are categorised in the 
Bogoliubov-de Gennes (BdG) classes, which comprise class D (p-wave 
SC), C (d-wave SC), DIII (p-wave time-reversal-symmetric SC), 
and CI (d-wave time-reversal-symmetric SC).  The $d+id$ SC 
belongs to class C.

In the context of the present article, a question 
is: In the Floquet $(d+id)$ pairing, whether and how 
would the flatband physics 
come in?  If we have a closer look at the amplitudes of the 
two-step correlated hopping $\Gamma$ and 
chiral spin coupling $J_{\chi}$, they behave\cite{kitamuraAoki22} as 
\begin{eqnarray}
\Gamma &\sim& t t' E^4 \times ({\rm function \, of }\;\omega, U) \nonumber \ \\
J_{\chi} &\sim& (t t')^2 E^4 \times ({\rm function \, of }\;\omega, U).
\end{eqnarray}
This directly shows that both terms increase with the 
second-neighbour hopping $t'$, 
i.e., increase with the degree of frustration.  In this sense, a 
situation closer to flatbands will promote 
the topological SC. 
The above is in line with the theoretical suggenstions 
for equilibrium that an existence of flat parts 
(such as van Hove singularities at which the group 
velocity vanishes) favours emergence of topological 
SC in equilibrium\cite{liu18}. 
It will be an interesting future problem how 
the laser-induced topological SC appears in 
(not partially but entirely) flatband systems.  

Now, an experimental question you might pose is: can we 
have intense enough laser for these topological phases?  
We can in general conceive a 
phase diagram for optical control of 
condensed matters on a parameter plane spanned by the laser's 
frequency $\omega$ and the light-field intensity $E$.  
Roughly, a line called ``Keldysh line" separates the 
plane into upper left and lower right halves.  In 
the Floquet engineering of electrons, we work in the former region, 
while the latter basically refers to optical properties\cite{aoki_RMP14}. 
On this plot, the laser used in McIver et al's experiment\cite{McIver2019} 
for FTI belongs to the former region, and the 
$\omega$-$E$ region required for the 
Floquet topological SC for typical parameters ($t, U, a$) 
of the high-Tc cuprates sits close to the region 
used by McIver et al.  
The required intensity will be reduced by 
going to lower $\omega$, but also by going to 
more frustrated lattices as we have just described.
Thus a final message of this section is that 
the flatband physics works 
both in one-body problems and in strongly-correlated systems.

\clearpage

\section{Other topics and outlook}

\begin{figure}[ht]
\begin{center}
  \includegraphics[width=0.55\textwidth]{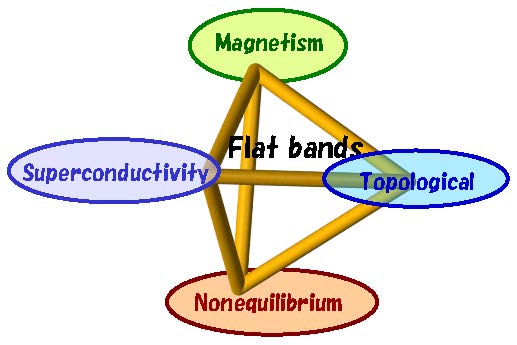}
\caption{Fields that are encompassed by the flatband physics.}
\label{fig:4pillars}     
\end{center}  
\end{figure}

As we have described, flatbands 
give a widened horizon of the 
condensed-matter physics, covering the 
fields as symbolised in 
{\bf Fig.}\ref{fig:4pillars}
.  The flatband physics is still extending its horizon, 
ranging from a guiding principle for materials design to 
concepts such as quantum metric. 
In searching for quantum phases, we have to examine competitions between 
superconductivity, SDW, CDW etc.  In the flaband SC, 
the nesting physics is irrelevant, which helps since 
other orders will not compete with SC in nesting-related ways.

There are a lot of subjects that cannot be covered 
here, which we touch upon here:

\par
\ \\

{\bf Boson systems and Bose-Einstein condensation on flatbands} 

\begin{figure}[ht]
\begin{center}
  \includegraphics[width=0.95\textwidth]{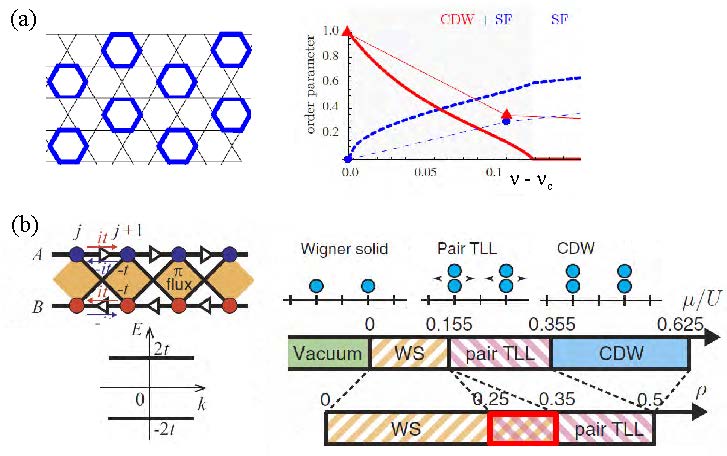}
\caption{(a) Left: An exact eigenstate of  the Bose-Hubbard model 
on Kagome lattice at filling $\nu = \nu_c = 1/9$, where each blue hexagon is the localised 
state on the flatband.  Right: Phase diagram against $\nu-\nu_c$, where curves 
represent the mean-field result, symbols an exact-diagonalisation result. [S.D. Huber and E. 
Altman, Phys. Rev. B 82, 184502 (2010).]  (b) Left: Bose-Hubbard model 
on Creutz ladder with $\pi$ flux, where the hopping is $it (-it)$ 
along (against) the direction of the arrow.  Inset depicts the band structure.  
Right: Phase diagram as a function of the
chemical potential $\mu$ or the density $\rho$ of bosons, obtained with exact diagonalisation and 
the Bethe ansatz solution.  Here WS: Wigner solid, 
TLL: Tomonaga-Luttinger liquid, CDW: charge-density wave, as schematically shown 
in the top inset.   [S. Takayoshi et al, Phys. Rev. A 88, 063613 (2013).]
}
\label{fig_supersolid}     
\end{center}  
\end{figure}

While we have 
concentrated on fermion systems on flatbands, 
boson systems on flatbands are also interesting.  
We are then talking about Bose-Einstein 
condensation in {\it bose-Hubbard model on flatbands}.  
One interest there is an emergence of ``supersolids" 
where superfluidity coexists with crystalline orders, 
see {\bf Fig.}\ref{fig_supersolid}.  
This was discussed by Huber and Altman\cite{huber10} 
for kagome and triangle-chain lattices, and also 
by Takayoshi et al\cite{takayoshi_PRA13} for the Creutz ladder.  In the 
latter, exact diagonalisation and Bethe ansatz solution 
are used to show the presence of a pair-Tomonaga-Luttinger 
liquid coexisting with Wigner solid in a phase diagram.

\par
\ \\

{\bf Field-theoretic view}

\begin{figure}[ht]
\begin{center}
  \includegraphics[width=0.6\textwidth]{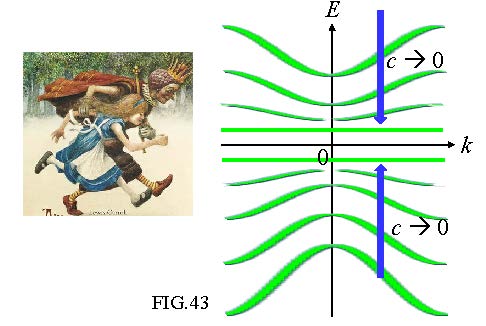}
\caption{Energy dispersion (Dirac field with electron and hole branches in the field theory, or 
valence and conduction bands in condensed matter) for $c \rightarrow 0$ is schematically 
shown, where $c$ stands for the speed of light in field theory, or Fermi velocity in condensed matter.  In this limit, what is called Carrollian symmetry emerges.  
Attached picture is from Lewis Carroll: {\it Alice in Wonderland}. 
}
\label{fig_Carroll}     
\end{center}  
\end{figure}

Flatbands have, deservingly, notable field-theoretic implications.  
This is treated in terms of what is called the Carroll symmetry.  
The background is the following: 
Poincar\'{e} algebra has given a basis for Einstein's theory of special relativity as well-known.
There, $c \rightarrow \infty$ limit (with $c$ being the speed of light) 
is called the ultra-relativistic limit (see e.g. Landau-Lifshitz: {\it Classical Theory of Fields}).  
What about the opposite limit of $c \rightarrow 0$, then 
({\bf Fig.}\ref{fig_Carroll})?   
$c \rightarrow 0$ algebra is known as ``Carrollian algebra" as a special case of Poincar\'{e} algebra, but has been considered just as a mathematical 
curiosity.  
(Incidentally, the nomenclature derives from 
Lewis Carroll of Alice in Wonderland.)  
In recent years, there is a surge of interests, especially in the conformal Carrollian algebra as a potential holographic dual of asymptotically flat 
spacetimes\cite{bagchi23}.  
In the $c \rightarrow 0$ limit, the Poincar\'{e} symmetry 
turns into the Carroll symmetry where only the time derivative survives 
with the spatial one disappearing, which makes the temporal evolution 
and spatial translation separated, resulting in an infinite 
number of symmetry generators (called supertranslations). 
Interdisciplinary links are 
to condensed-matter systems 
(the flatbands of course), as well as to cosmology, etc, where 
strange phenomena abound.  
Thus, contrary to a naive expectation that a particle simply 
would become immobile for $c \rightarrow 0$, 
a nontrivial dynamics can emerge in the field theory.  
This strongly reminds us of the fact that the flatband is certainly distinct from 
a trivial atomic limit ($t \rightarrow 0$ in the TB model) 
as we have stressed in terms of the strange Hilbert space.  
Another condensed-matter topic where field theory is 
applied in a standard way is the localisation 
problem,  and  disordered flatband systems could be 
intriguing\cite{hatsugai_localization21}.

%

Other topics include flatbands in photonic bands, 
see, e.g., Ref.\cite{endo}.  
Since the flatbands incorporate peculiar 
quantum metric properties, implications for 
quantum informations may be anticipated as 
well.  
Let us finish by saying we can anticipate 
further developments into diverse directions, including 
the interdisciplinary ones.  
\par
\ \\

\clearpage

{\bf Acknowledgements}: The author wishes to thank 
Kazuhiko Kuroki, P\"{a}ivi T\"{o}rm\"{a}, 
Yasuhiro Hatsugai, Sharareh Sayyad, Sota Kitamura, 
Philipp Werner, Yoshiro Takahashi, Bohm-Jung Yang, Tomoki Ozawa, Izumi Hase, Tatsuhiro Misumi 
and Hiroyuki Tajima 
for invaluable discussions on the flatband physics over the years.

\end{document}